\documentclass[12pt]{article}
\usepackage[utf8]{inputenc}
\usepackage{setspace, natbib, amsthm, amsfonts, amssymb, amsmath, mathrsfs, rotating, epsfig, booktabs, pdflscape, subfig, tabularx, titlesec, bezier, har2nat, color, comment, tikz, pgfplots, xfrac, bm, esdiff, mathtools, indentfirst, nicefrac, fullpage, threeparttable}
\usepackage{comment}
\usepackage{amssymb,amsfonts,bbm,amsthm,amsmath,verbatim,graphicx}
\usepackage[utf8]{inputenc}
\usepackage{xparse}
\usepackage{tikz}
\usepackage[margin=1in]{geometry}
\usepackage{setspace}
\usepackage{soul}
\usepackage{comment}

\usepackage{enumitem} 

\emergencystretch=\maxdimen
\usetikzlibrary{matrix,backgrounds}
\pgfdeclarelayer{myback}
\pgfsetlayers{myback,background,main}

\tikzset{mycolor/.style = {line width=1bp,color=#1}}%
\tikzset{myfillcolor/.style = {draw,fill=#1}}%

\NewDocumentCommand{\highlight}{O{blue!40} m m}{%
\draw[mycolor=#1] (#2.north west)rectangle (#3.south east);
}

\NewDocumentCommand{\fhighlight}{O{blue!40} m m}{%
\draw[myfillcolor=#1] (#2.north west)rectangle (#3.south east);
}

\newcommand{\R}{\mathbb{R}}

\newcommand{\E}{\mathbbm{E}}

\usepackage{hyperref}

\title{Forecasting Algorithms for Causal\\\ Inference with Panel Data\thanks{For helpful comments, we are grateful to Guido Imbens, Aniket Kesari, Lihua Lei, Jann Spiess, and seminar participants at the ALEA and CELS Annual Conferences and the NBER Summer Institute Forecasting and Empirical Methods Meeting. An implementation of the SyNBEATS estimator is available at \url{https://github.com/Crabtain959/SyNBEATS}.}}
\author{
    Jacob Goldin\thanks{University of Chicago and NBER, email: jsgoldin@uchicago.edu}
    \and
    Julian Nyarko\thanks{Stanford University. Corresponding Author, email: jnyarko@law.stanford.edu}\\
    \and
    Justin Young\thanks{Stanford University, email: justiny@stanford.edu}
}

\date{\today}

\begin{document}

\maketitle
    
\onehalfspacing

\begin{abstract}

Conducting causal inference with panel data is a core challenge in social science research. We adapt a deep neural architecture for time series forecasting (the N-BEATS algorithm) to more accurately impute the counterfactual evolution of a treated unit had treatment not occurred. Across a range of settings, the resulting estimator (``SyNBEATS'') significantly outperforms commonly employed methods (synthetic controls, two-way fixed effects), and attains comparable or more accurate performance compared to recently proposed methods (synthetic difference-in-differences, matrix completion). An implementation of this estimator is available for public use. Our results highlight how advances in the forecasting literature can be harnessed to improve causal inference in panel data settings.

    
\end{abstract}
\newpage

\section{Introduction}
Conducting causal inference with panel data is a core challenge in social science research. Consider a panel of states, one of which (the treated state) adopts a new policy. What is the effect of this policy on some outcome of interest? The question can be cast as a prediction problem: the (potential) untreated outcomes for the treated state can be observed in the time periods prior to the new policy's adoption, but not afterwards. Meanwhile, the potential untreated outcomes for the control states can be observed both before and after the new policy's adoption. Identifying the causal effect of the policy entails estimating what the outcome would have been in the treated state during time periods after the policy's adoption, had the new policy not been adopted \cite{holland1986}. 
Alternative tools for causal inference in panel data settings can be understood as different methods for imputing these counterfactual outcomes on the basis of the data that is observed: namely, data from the treated state prior to the policy's adoption, and from the control states in time periods both before and after the policy's adoption.

In this paper, we draw on new advances in the time series forecasting literature to improve the accuracy of the imputations employed for causal inference in panel data settings. Over the past few years, the forecasting literature has proposed a number of deep neural architectures that have significantly improved predictive capabilities over older models. However, these models are designed to be applied to data for single time series, i.e., predicting future values of a unit based on that same unit's past values. For causal inference with panel data, this is an important limitation because single-unit time series models do not incorporate information from the time series of control unit outcomes to help estimate missing values for the treated unit. We overcome this limitation by incorporating the time series of outcomes for control units into the forecasting model for the treated unit as additional features. That is, to impute the potential untreated outcome in the treated unit following the treatment, we feed into the model the outcome from the control states during the same time period -- effectively casting contemporaneous outcomes of the control states as ``leading indicators'' for the potential untreated outcome of the treated state. 

Among the different deep architectures for forecasting models, we focus on the neural basis expansion analysis for time series (N-BEATS) algorithm \cite{oreshkin2019}. N-BEATS is a deep neural architecture designed to predict future values in a time series on the basis of past values. The algorithm has been shown to perform well in a range of forecasting tasks. The key innovation we rely on is a recent adaptation of N-BEATS that incorporates time series other than the one being predicted as additional features \cite{olivares2019}; in the panel data setting, this innovation allows us to use N-BEATS to predict unobserved values of the treated unit on the basis of prior values of the treated unit as well as contemporaneous values of the control units. Because our proposed approach essentially involves using the N-BEATS algorithm to estimate a synthetic (i.e., predicted) untreated outcome for the treated state during the post-treatment period, we refer to it as Synthetic N-BEATS (``SyNBEATS'').

Although the N-BEATS algorithm has been shown to excel at a range of forecasting tasks, an important concern is whether its performance will be as strong when applied to the relatively small panel data sets typically employed in social science research. With limited data, simpler methods like synthetic controls (SC) or two-way fixed effects (TWFE) may yield more reliable causal estimates. To assess the suitability of SyNBEATS for causal inference with panel data, we compare it to existing alternatives across two canonical panel data settings. Specifically, we contrast performance in data that has been used to estimate the effect of a cigarette sales tax in California \cite{abadie2010} and the German reunification on the West Germany economy \cite{abadie2015}. In both of these settings, we find that SyNBEATS outperforms canonical methods such as SC and TWFE estimation. In addition, we compare the performance of these models in estimating the impact of simulated events on abnormal returns in publicly traded firms \cite{baker2020machine}. In this setting, where historical values would not be expected to provide much information about future outcomes, SyNBEATS only marginally improves performance relative to the other estimators. We also compare SyNBEATS to two recent proposed causal inference methods for panel data settings: matrix completion (MC) \cite{mcathey2021} and synthetic difference-in-differences (SDID) \cite{sdid}. In the three settings we consider, we find that SyNBEATS generally achieves comparable or better performance compared to synthetic difference-in-differences, and significantly outperforms matrix completion. We further investigate the factors that shape the relative performance of SDID and SyNBEATS through a range of simulations. Finally, we unpack SyNBEATS' strong comparative performance, and find it stems from both model architecture and SyNBEATS' efficient use of time-series data in informing its predictions.

Our findings build on a growing literature applying machine learning tools to casual inference. In addition to the new methods discussed above, two recent papers similar in spirit to ours include \citet{muhlbach2021tree}, which adapts tree-based methods to synthetic control estimators, and \citet{poulos2021rnn}, which applies a recurrent neural network model to predict treated units' outcomes based on pre-treatment observations. As discussed below, our approach differs from these in that SyNBEATS forms its counterfactual prediction directly using both pre-treatment values of the treated unit 
and post-treatment values of the untreated units. 

The remainder of the paper proceeds as follows: Section \ref{sec:SyNBEATS:ex} describes SyNBEATS and related estimators. Sections \ref{sec:exercises} and \ref{sec:recent} compare SyNBEATS to common and recent estimators, respectively. Section \ref{sec:gains} explores the source of SyNBEATS' performance. Section \ref{sec:conclusion} concludes.

\section{Theoretical Framework
\label{sec:SyNBEATS:ex}}

This section describes our setup, introduces SyNBEATS, and describes alternative panel data estimators.

\subsection{Setup}

\begin{figure}

\centering
\subfloat[Problem Statement]{\label{main:a}\hspace{0cm}\includegraphics[scale=.3]{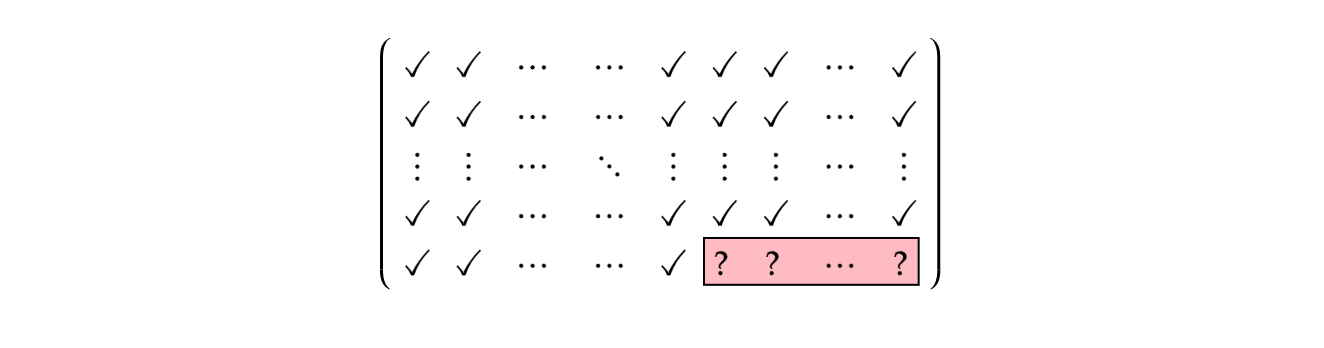}}\par\medskip
\begin{minipage}{.4\linewidth}
\centering
\subfloat[``Vertical'' Estimation]{\label{main:b}\includegraphics[scale=.3]{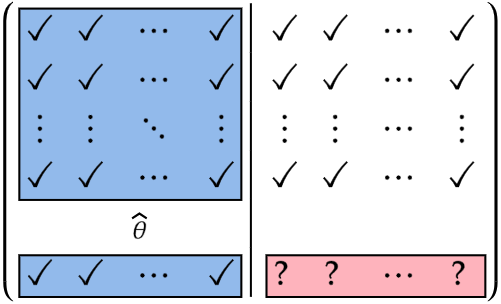}}
\end{minipage}%
\begin{minipage}{.4\linewidth}
\centering
\subfloat[``Horizontal'' Estimation]{\label{main:c}\includegraphics[scale=0.3]{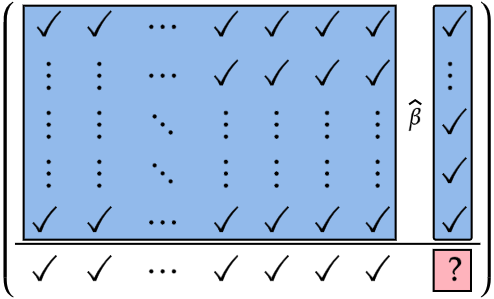}}
\end{minipage}

\caption{Problem statement and common estimation approaches. Panel (a) illustrates the core challenge of imputing the unobserved counterfactual for the treated unit. Panels (b) and (c) illustrate the respective use of information about contemporaneous outcomes for other units (vertical) and lagged outcomes for the same unit (horizontal).}
\label{fig:intro_problem}
\end{figure}

We consider a panel data setting with $N$ units across $T$ time periods. One unit ($N$) adopts a new policy that takes effect for every period after $T_0<T$. We use $W_{it} \in \{0,1\}$ to indicate whether a unit has adopted the policy, so that $W_{it} = 1$ for $i=N$ and $t> T_0$, and $W_{it} = 0$ otherwise. We are interested in the causal effect of the policy on some outcome of interest $Y_{it}$. Let $Y_{it}(1)$ and $Y_{it}(0)$ denote the (potential) outcomes of interest for unit $i$ in period $t$ corresponding to $W_{it}=1$ and $W_{it}=0$, respectively. For any unit and time period, the researcher observes only one potential outcome: $$Y_{it} = W_{it} Y_{it}(1)  + (1-W_{it})  Y_{it}(0)$$  
The econometric challenge, illustrated in Figure \ref{fig:intro_problem}(a), is to impute the missing counterfactual outcomes, $Y_{Nt}(0)$ for $t>T_0$, in order to estimate the average treatment effect on the treated: $$\tau^{ATT}= E[Y_{it}(1) - Y_{it}(0) | W_{it} = 1]$$

\subsection{Approaches to Panel Data  Estimation} 
Many methods have been proposed to conduct causal inference in panel data settings. 
One popular approach is the synthetic control (SC) estimator \cite{abadie2003, abadie2010}
, which constructs a synthetic counterfactual for the treated unit in the post-treatment periods by finding weights on relevant control unit values such that the synthetic unit fits the treated unit values well in the pre-treatment period. 
\citet{doudchenkoimbens2016} expand upon the SC literature and show these methods can be viewed as leveraging the structure of the data \textit{vertically} across units (Figure \ref{fig:intro_problem}(b)). In particular, the weights on the control units can be interpreted as being derived from a regression of treated unit outcomes on control outcomes in the pre-treatment period: \begin{equation*} \hat{\theta} = \text{arg min}_\theta \sum_{t\leq T_0} \left( Y_{Nt} - \theta_0 - \sum_{i=1}^{N-1} \theta_i Y_{it} \right)^2  \label{syntheticcontrol} \end{equation*}  where the weights are non-negative and sum to 1, and $\theta_0$ is typically constrained to be 0. 

Although less common in panel data estimation, another approach is to assume
that temporal patterns are stable across units to exploit the structure of the data \textit{horizontally} (Figure \ref{fig:intro_problem}(c)). In the panel data setting, this ``unconfoundedness'' approach
extrapolates treated unit counterfactual values in $t>T_0$ by using weights obtained from regressing the control outcomes in $t>T_0$ on those in prior periods. 
Formally, with the number of treated units indexed by $\mathcal{I}$, we can view the weights as the solution to the following minimization problem: \begin{equation*} \hat{\beta} = \text{arg min}_{\beta} \sum_{i \notin \mathcal{I}} \left( Y_{iT} - \beta_0 - \sum_{s=1}^{T-1} \beta_s Y_{is} \right)^2 \label{unconfoundedness} \end{equation*}

Whereas the methods discussed thus far rely exclusively on either vertical or horizontal information, two-way fixed effects (TWFE) estimators  simultaneously exploit stable patterns over time and across units. In the simplest case with no covariates, the TWFE estimate for the treatment effect is obtained from: 
\begin{equation*}
    (\hat{\tau}^{TWFE}, \hat{\mu},\hat{\alpha},\hat{\delta}) = \text{arg min}_{\tau,\mu,\alpha,\delta} \left\{ \sum_{s=1}^N\sum_{t=1}^T ( Y_{st} - \mu -\alpha_s - \delta_t -  \mathbf{1}_{\{s=N\}} \mathbf{1}_{\{t > T_0 \}} \tau)^2 \right\}
\end{equation*} 

Interactive fixed effects and more generally factor models extend this literature to allow for richer unobserved heterogeneity \cite{bai2002, bai2003}. 
These more general models assume the data generating process is given by a linear function of observed covariates and an unobserved low-rank matrix plus noise, allowing for arbitrary interactions beyond simple additive effects. Without covariates and with binary treatment, the outcomes can be described as $$\boldsymbol{Y} = \boldsymbol{L} + \boldsymbol{W} \odot \boldsymbol{\tau} + \boldsymbol{\epsilon}$$ where $\boldsymbol{L}$ represents the target estimand we wish to recover. In the factor model literature, $ \boldsymbol{L}:= \boldsymbol{UV}^T$ with loadings $\boldsymbol{U}$ and factors $\boldsymbol{V}$. 


A recent estimator  for $\boldsymbol{L}$ proposed by \citet{mcathey2021} draws on the matrix completion (MC) literature. MC aims to recover missing entries in a data matrix that are missing at random, in contrast to the ``block'' missingness present in panel data causal inference settings (Figure 1(a)). 
\citet{mcathey2021} incorporate the intuition of TWFE into matrix completion by solving the standard low-rank matrix completion problem but keeping fixed effects explicitly unregularized. 

Another recent method simultaneously incorporating both vertical and horizontal information is the synthetic difference-in-difference (SDID) estimator. 
First, it employs a SC-type approach to reweight the unexposed control units, creating a parallel trend. 
Second, SDID applies a difference-in-differences type analysis on this reweighted panel. In contrast to standard difference-in-differences, SDID allows time period weights to vary to accommodate the possibility that more recent periods are more informative. 
With both sets of weights, the estimator recovers the treatment effect by solving: $$(\hat{\tau}^{SDID}, \hat{\mu},\hat{\alpha},\hat{\delta}) = \text{arg min}_{\tau,\mu,\alpha,\delta} \left\{ \sum_{s=1}^N\sum_{t=1}^T ( Y_{st} - \mu -\alpha_s - \delta_t -  \mathbf{1}_{\{s=N\}} \mathbf{1}_{\{t > T_0 \}} \tau)^2 \hat{\omega}_i \hat{\lambda}_t \right\}$$ 


\subsection{SyNBEATS} 

\begin{figure}[h!]
    \centering
    \includegraphics[scale=0.35]{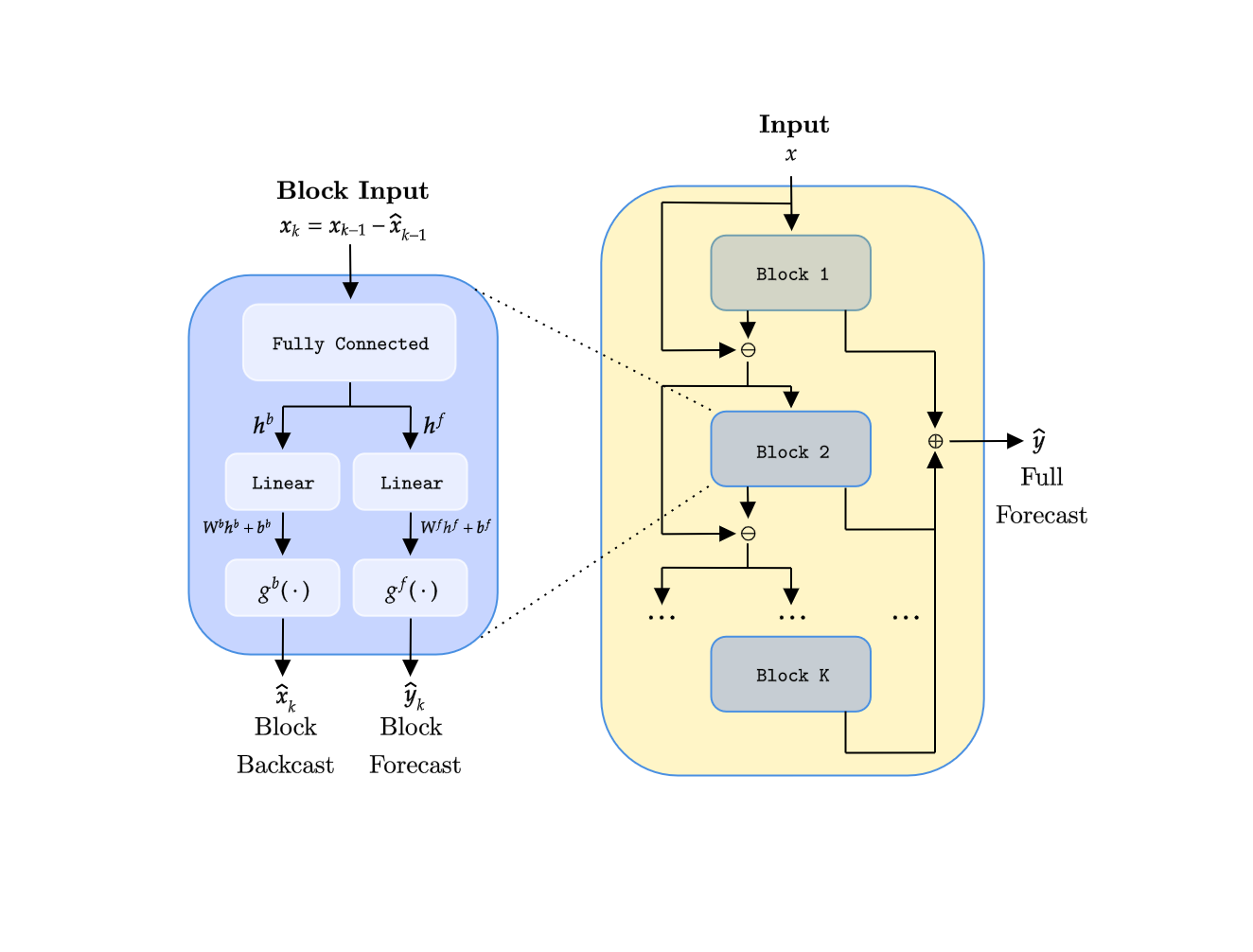}
    \caption{Model Architecture. \textit{Left}: A single block $k$. \textit{Right:} The full model.}
    \label{fig:SyNBEATSarchitecture}
\end{figure}

We cast the challenge of causal inference in the panel data setting as a supervised learning problem and propose drawing on both horizontal and vertical information to inform the prediction of the (missing) untreated potential outcomes for the treated unit in the post-treatment time period. Specifically, we propose a neural network architecture to determine prediction weights across time and units, depicted in Figure \ref{fig:SyNBEATSarchitecture}.  

Model training is based on the mean squared error loss over the pre-intervention period when $Y_{Nt}$ is fully observed. To form predictions for these values at every time step, the model uses the contemporaneous control outcomes at $t$ and the lagged treated unit's outcomes from $t-n_{\text{lag}}$ to $t-1$, where $n_{\text{lag}}$ denotes the number of lagged outcomes used. Because $n_{\text{lag}}$ is constant across all training examples, we define $\underline{t} := t-n_{\text{lag}}$ for convenience. 
Formally, at every $t = 1,\ldots,T_0$, our model predicts the observed $Y_{Nt}$ with $\{Y_{it} \}_{i=1}^{N-1}$ and $\{Y_{Ns}\}_{s=\underline{t}}^{t-1}$. 
The model, $m: \R^{(N-1)+(n_\text{lag})} \to \R$, is trained via gradient descent methods to minimize loss based on the following loss function:\footnote{The loss function assumes a one-period-ahead prediction objective, which is what we implement in practice. The method can also support multiple-period-ahead predictions by averaging over the loss in subsequent future periods.} 


\begin{equation*}
    J(\theta) =   \frac{1}{T_0}\sum_{t \leq T_0} \left(Y_{Nt} - m \left( \{Y_{it} \}_{i=1}^{N-1}, \{Y_{Ns}\}_{s=\underline{t}}^{t-1} \ ; \ \theta \right) \right)^2 
\end{equation*}

In a standard feedforward neural network, each training input $\boldsymbol{x} \in \R^{d_0}$ is fed through multiple \textit{layers}, with each layer $\ell$ generating an intermediate output $\boldsymbol{h}^{[\ell]} \in \R^{d_\ell}$.\footnote{For ease of notation, $\ell = 1,\ldots,L$ denotes which layer is of interest.} This relationship is governed by $$\boldsymbol{h}^{[\ell]} = f^{[\ell]}(\boldsymbol{W}^{[\ell]}\boldsymbol{h}^{[\ell-1]} + \boldsymbol{b}^{[\ell]})$$  Here, $\boldsymbol{W}^{[\ell]} \in \R^{d_\ell \times d_{\ell-1}}$ and $\boldsymbol{b}^{[\ell]} \in \R^{d_\ell} $ represent weight and bias matrices akin to standard linear regression. In the first layer, $\boldsymbol{h}^{[0]} \equiv \boldsymbol{x}$. The function $f(\cdot)$ is taken here to be $f(x) = \max\{0,x\}$, so that $f$ explicitly introduces non-linearity into the network to allow it to learn non-linear relationships in the data. After an arbitrary number of layers $L$ with intermediate outputs $\boldsymbol{h}^{[1]}, \ldots, \boldsymbol{h}^{[L]}$, the model yields a final output $\boldsymbol{\hat{y}} \in \R^{d_y}$. This in turn is compared to the observed outcome $\boldsymbol{y} \in \R^{d_y}$ and a standard loss function (e.g. MSE) is used to compute the overall loss. As each $\boldsymbol{h}^{[\ell]}$ depends on all components of $\boldsymbol{h}^{[\ell-1]}$ at each layer, we call such a network \textit{fully connected}. Consistent with other nonparametric estimation methods, a closed-form solution does not exist generically, and the parameters must be estimated via gradient descent methods. In neural networks, this numerical method proceeds in a familiar manner: the gradient of the loss function propagates backward through the network via the chain rule to update all weights $\boldsymbol{W}^{[\ell]}$ and $\boldsymbol{b}^{[\ell]}$ until the weights converge.\footnote{Without any assumptions or additional structure on the problem, a global minimum is not guaranteed, but in practice even local minima may perform well in RMSE compared to standard econometric models.}


In contrast to a standard feed-forward neural network, the N-BEATS architecture is compartmentalized into different \textit{blocks}. Given a time series input $\boldsymbol{x} \in \R^{d_0}$, where $d_0$ is the length of the lookback period used to forecast, each block $k$ adopts this standard neural network structure but generates not only a forecast $\boldsymbol{\hat{y}} \in \R^{d_y}$ with $d_y$ as the length of the forecast horizon, but also a {backcast} $\boldsymbol{\hat{x}} \in \R^{d_0}$. In its original implementation, $\boldsymbol{x}$ is composed of historical values of the time-series being predicted (e.g., lagged outcomes of the treated unit). \citet{olivares2019} modifies this approach to allow for predictions using information from additional (covariate) time series.  As shown in Figure \ref{fig:SyNBEATSarchitecture}, the model makes its prediction in each block via a set of shared fully connected layers that branch into two separate forecast and backcast layers. 
For each block $k$, the shared fully connected layers yield intermediate outputs $\boldsymbol{h}^{[1]}_k,\ldots, \boldsymbol{h}^{[L]}_k$ as before. Next, $\boldsymbol{h}^{[L]}_k$ is split into $\boldsymbol{h}^f_k$ and $\boldsymbol{{h}}^b_k$, which serve as the final intermediate outputs for the forecast and backcast, respectively. The predictions in each block are finally given by:  \begin{align*}
    \hat{\boldsymbol{y}}_k & = g^f_k ( \boldsymbol{W}^f_k \boldsymbol{h}^f_k + \boldsymbol{b}^f ) \\ 
    \hat{\boldsymbol{x}}_k & = g^b_k (\boldsymbol{W}^b_k  \boldsymbol{h}^b_k + \boldsymbol{b}^b ) 
\end{align*}
where $g^f(\cdot)$ and $g^b(\cdot)$ are linear projection operators.

The backcast at block $k$, $\boldsymbol{\hat{x}}_{k}$, is fed to the next block $k+1$ as the new input: \begin{align*} \boldsymbol{x}_k & =  \boldsymbol{x}_{k-1} -  \boldsymbol{\hat{x}}_{k-1} \end{align*} The forecast at block $k$, $\boldsymbol{\hat{y}}_{k}$, is added iteratively to the forecasts at all other previous blocks: \begin{align*} \boldsymbol{\hat{y}}_{\text{block}} & = \sum_k \boldsymbol{\hat{y}}_k\end{align*} In other words, this block structure breaks down the signal from input $\boldsymbol{x}$ step-by-step, with each $\boldsymbol{\hat{x}}_k$ representing the signal leftover that could not be explained by the previous neural networks in blocks $1, \ldots, k$. Although in theory an arbitrarily large, standard feed-forward neural network should be able to flexibly mine all input signal and achieve arbitrarily low error, in practice these alternatives struggle to do so within reasonable computational limits. 

SyNBEATS largely follows \citet{oreshkin2019} and \citet{olivares2019} in its model architecture. The sole modification is that we augment the feature set to include outcomes for contemporaneous control units, in addition to historical values of the treated unit (i.e., the time series being predicted). To do so, we concatenate the contemporaneous controls and the lagged outcomes into a single vector. Each training example at time step $t$ is thus represented as: $$\boldsymbol{x}_t := [\{Y_{it}\}_{i=1}^{N-1}, \{Y_{Ns}\}_{s=\underline{t}}^{t-1}]$$ This is then is fed through the model architecture to predict $\boldsymbol{y}_t := [\{Y_{Ns}\}_{s=t}^{\overline{t}}]$, where $\overline{t}-t+1$ refers to the number of future periods being predicted. 

In our implementation of SyNBEATS, we set $\overline{t}=t$. To predict beyond a single post-treatment period, we iteratively use the predicted value of the preceding period as an input to the algorithm; for example, in order to predict $\widehat{Y}(0)_{s,t+1}$, we use $\widehat{Y}(0)_{st}$ as an input. 

\subsection{Identification}


As described above, SyNBEATS entails training a supervised model to predict the untreated potential outcome for the treated unit based on prior (untreated) values of the treated unit and contemporaneous values of the control units:
\begin{align*}
    \hat{Y}_{Nt}(0) =  m \left( \{Y_{it}(0) \}_{i=1}^{N-1}, \{Y_{Ns}(0)\}_{s=\underline{t}}^{t-1}\right)
\end{align*}
For SyNBEATS to accurately estimate the average treatment effect on the treated, $\tau$, the relationship between the label, $Y_{Nt}$, and the features,  $\{Y_{it}(0) \}_{i=1}^{N-1}, \{Y_{Ns}(0)\}_{s=\underline{t}}^{t-1}$ during the post-period must be learnable from data observed during the pre-period. A sufficient condition for this to hold is that the distribution of outcomes is stationary and the timing of treatment is exogenous. Similar conditions are commonly imposed to justify the use of synthetic controls and related estimators.  



That being said, due to the neural network architecture underlying the estimator, it is difficult to provide asymptotic guarantees on SyNBEATS's behavior. Indeed, highly flexible estimators are generally biased due to overfitting and regularization; the same is true for commonly employed estimators such as synthetic controls. In some applications, researchers nonetheless may be particularly interested in the consistency of the estimator.  One method for obtaining a consistent SyNBEATS estimator is via bias-correction \cite{ttest, viviano}. Under this approach, the researcher evaluates the estimator's bias using a subset of the pre-treatment sample. This can then be subtracted from the SyNBEATS-estimated treatment effect to yield an asymptotically consistent estimate. Under standard event study assumptions (exogeneity of treatment timing, stationarity of outcomes, etc.), we can adapt the proof from \citet{viviano} to show that this bias-adjusted version of SyNBEATS yields a consistent treatment effect estimate. We develop this estimator and compare its performance to our baseline SyNBEATS estimator in Section \ref{sec:gains} below.

\section{Comparison of SyNBEATS to Popular Estimators}\label{sec:exercises}
In this section, we compare the performance of SyNBEATS to two popular methods for causal inference in panel settings: synthetic controls and two-way fixed effects estimation. We analyze these methods in three data settings.

\subsection{Framework}

In each setting, we collect outcome data $Y_{it}$ for a balanced panel of $i\in\{1,2,...,N\}$ untreated units and $t\in\{1,2,..,T_0, T_0+1, ..., T\}$ periods. 
As discussed in Section 2.1, the econometric challenge is to estimate the counterfactual untreated outcome $Y_{Nt}(0)$ for unit $N$ when $t>T_0$. We compare alternative estimators for this term by analyzing the prediction error. To do so, we conduct placebo exercises assuming $Y_{Nt}(0)$ is unobserved (by masking actually observed outcomes) for $t > T_0$ and predicting $Y_{Nt}(0)$ (out-of-sample) using other data from the panel that would be available to the researcher if the unit really had been treated (i.e. if $Y_{Nt}(0)$ really was unobserved). We consider two types of validation exercises, involving: (1) the prediction error obtained from untreated control units in the post-treatment period, and (2) the prediction error for the treated unit during the pre-treatment period. By comparing estimates of $Y_{Nt}(0)$ derived from alternative methods to the true $Y_{Nt}(0)$, we can evaluate the performance of the various methods in estimating $\tau^{ATT}$.

\subsection{Implementation of Estimators}

This subsection briefly discusses our implementation of SyNBEATS, SC, and TWFE.

\subparagraph{Synthetic Controls} We implement the SC method as proposed by \citet{abadie2010}. As recommended in the subsequent literature, we use pre-treatment outcomes to assess a control state's suitability for inclusion in the SC \cite{cavallo2013, doudchenkoimbens2016}.

\subparagraph{Two-Way Fixed Effects}
We consider a standard TWFE regression model without time-varying covariates. For consistency with the other methods we consider, we allow the treatment effect to vary by post-treatment year, and estimate the following model:
\begin{equation*}
    Y_{st} = \alpha_s + \delta_t + \sum_{i=T}^{T+K} \tau_i \mathbf{1}_{\{s=N\}} \mathbf{1}_{\{t=i\}} + \varepsilon_{st} 
\end{equation*}
Unlike the other approaches considered here, we estimate the TWFE model on both the pre-treatment values of $Y_{st}(0)$ as well as the post-treatment values of $Y_{st}(1)$. The model's implied prediction of the potential untreated outcome in a given post-treatment year can be derived from the estimated treatment effect in the corresponding year:
\begin{equation*}
    \widehat{Y}_{st}(0) = Y_{st}(1) - \widehat{\tau}_{t}
\end{equation*}

\subparagraph{SyNBEATS} We implement our SyNBEATS algorithm as described in Section \ref{sec:SyNBEATS:ex}. We use the original hyperparameters as described by \citet{oreshkin2019}, as they have shown to yield high performance in a large number of time series settings and our applications have limited data available for hyperparameter tuning.\footnote{Specifically, our architecture includes 30 stacks, a single block with 4 layers and a layer width of 256. We use the Adam optimizer with default settings, including a learning rate of 0.001.} For the same reason, we set $n_{\text{lag}} = 1$. Although in principle $n_{\text{lag}}$ is a hyperparameter that can be tuned to improve performance, for convenience we treat it as fixed. 

\subsection{Proposition 99 in California}
In 1988, California implemented Proposition 99, raising the state tax on cigarettes from 10 cents to 35 cents per pack. \citet{abadie2010} apply the SC method to this setting to estimate the effect of the tax on cigarette sales. Comparing the real California to its synthetic counterpart, they find that the tax reduced cigarette sales in the years following its adoption. 

Below, we perform two exercises to assess the performance of SyNBEATS in this setting. The period of observation in the data set ranges from 1970 to 2000. The post-treatment period begins in 1989. Our outcome of interest is per capita cigarette sales (in packs), which we observe at the state-year level. Following \citet{abadie2010}, we take as our control group the 38 states that did not adopt tax increases during this period.

\subsubsection*{Exercise 1: Pseudo-Treated States}
In our first exercise, we exclude California (the true treated state). Instead, we assume that one other state from the control group (the ``pseudo-treated'' state) has been treated with a cigarette sales tax in 1989. We mask the post-1988 cigarette sales of the pseudo-treated state and apply each of the alternative causal inference methods. For each method, we assess the prediction error for the (actually untreated) outcome in the pseudo-treated state in the post-treatment period. This exercise will yield a valid assessment of the prediction errors for California in 1989 to the extent that, for each method we consider, the 1989 prediction error for California is drawn from the same distribution as the 1989 prediction error for the control states.

We repeat this process for the 38 control states, iteratively defining each control state as the pseudo-treated state. We first focus on the prediction error for the post-treatment year (1989). This yields 38 prediction errors for each estimation method -- one for each pseudo-treated state. To summarize the performance of a method, we compare the root mean squared error (RMSE) of the predictions across control states.

We also evaluate each model based on a longer-term, 5-year prediction window (1985--1989). In this case, each state will have five prediction errors, one for each post-treatment period. For the longer-term predictions, we calculate mean squared error based on the prediction errors in each pseudo-treated state over each post-treatment year.

\begin{figure} 
\centering
\subfloat[1-Year Predictions]{\includegraphics[width=.8\textwidth] {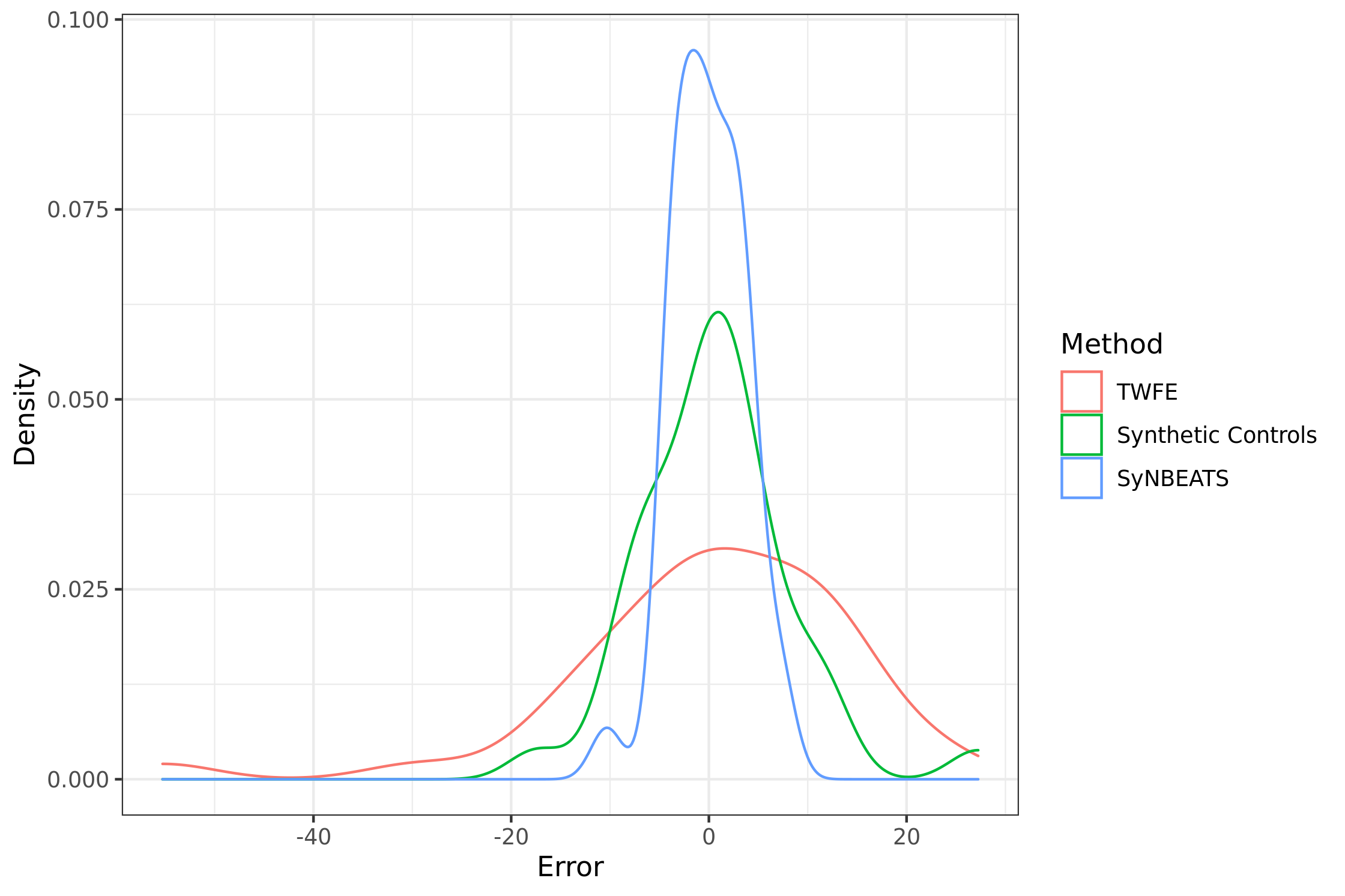}} \\
  \subfloat[5-Year Predictions]{\includegraphics[width=.8\textwidth] {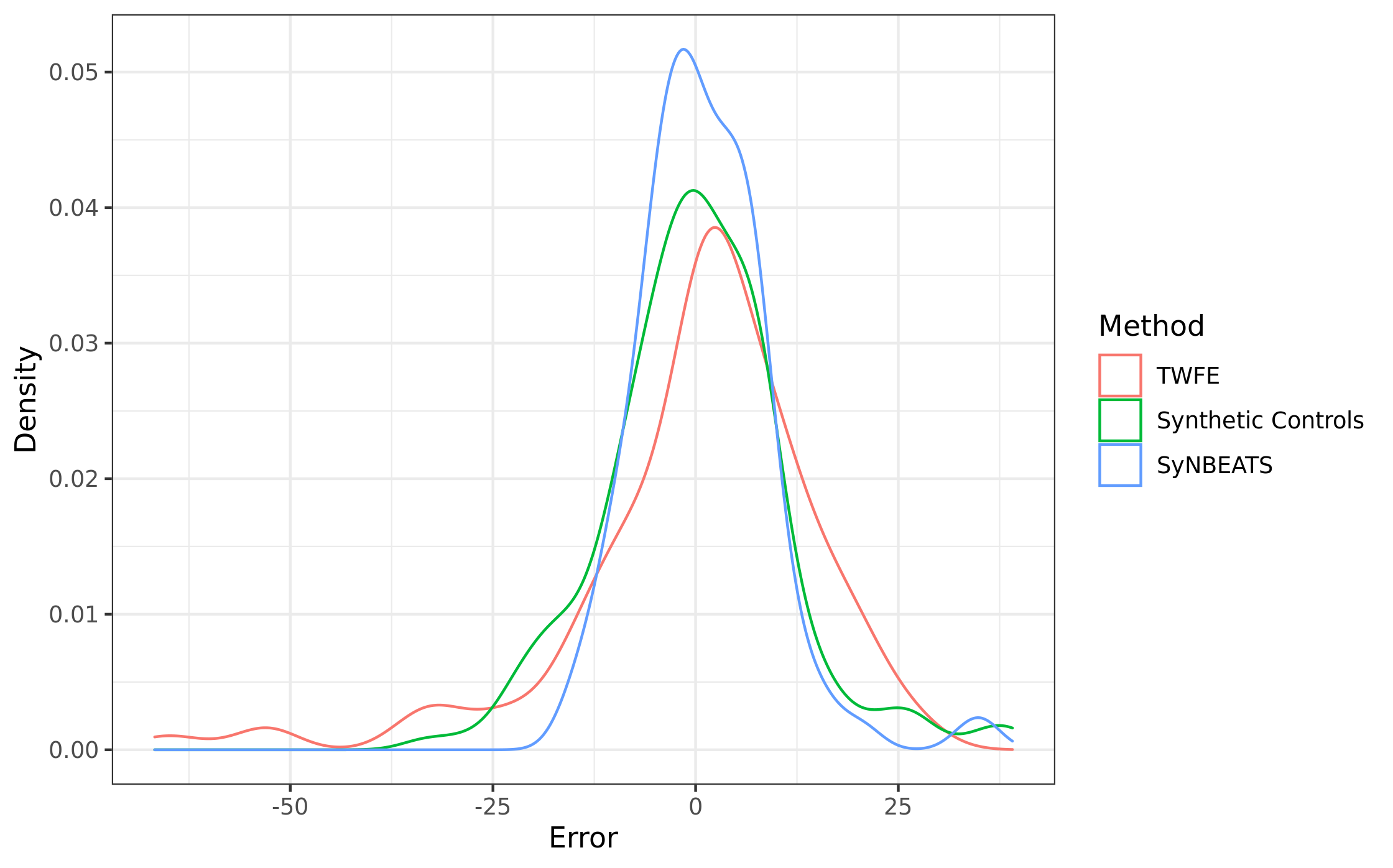}}
\caption{{Prop 99 Prediction Errors from Placebo Treated States}
\\\\\footnotesize The figure shows the distribution of prediction errors for pseudo-treated states, by prediction method, for the Proposition 99 data set. In Panel A, the outcome being predicted is the pseudo-treated state's per capita cigarette sales in 1989 ($N=38$); the model is trained on data from the pseudo-treated state from 1970-1988 and from the other control states for 1989. In Panel B, the outcome being predicted is the pseudo-treated state's per capita cigarette sales in years 1985--1989; the model is trained on data from the pseudo-treated state from 1970--1984 and from the other control states for the year being predicted.}
\label{f:smoking_states}
\end{figure}

Because the out-of-sample prediction error determines the accuracy of the estimated treatment effect, we compare the various estimation methods along this dimension. The model predictions are visualized in Figure \ref{f:smoking_states}, and the resulting distribution of prediction errors is summarized in Table \ref{t:traditional}. Among the methods we  consider, SyNBEATS yields the most accurate prediction, with an RMSE of 3.59, 54\% improvement over the second-best alternative of SC. Similarly, for longer-term predictions, SyNBEATS yields the best performance, with an RMSE of 8.17, which is a 27\% improvement over SC as the second-best alternative.

\subsubsection*{Exercise 2: Pseudo-Treated Years}
In this second exercise, we maintain California as the treated state, but we (counterfactually) assume that Proposition 99 took effect during some year prior to 1989 (the ``pseudo-treated year''). We iteratively define as our pseudo-treated year each year between 1975 and 1988. To assess longer-term predictions, we also consider 5-year pseudo-post-treatment periods in the same calendar year window (i.e., 1975--1979 through 1984--1988). In each iteration, we use data from years prior to the pseudo-treated year to estimate the predictive models.\footnote{The one exception is the TWFE model, which, as described above, we estimate using data that includes pseudo-treated years.} This exercise will yield a valid assessment of the prediction errors for California in 1989 to the extent that, for each method we consider, California's prediction error in 1989 is drawn from the same distribution as its prediction error for earlier years.

\begin{figure}[!ht]
\centering
\includegraphics[width=.8\textwidth] {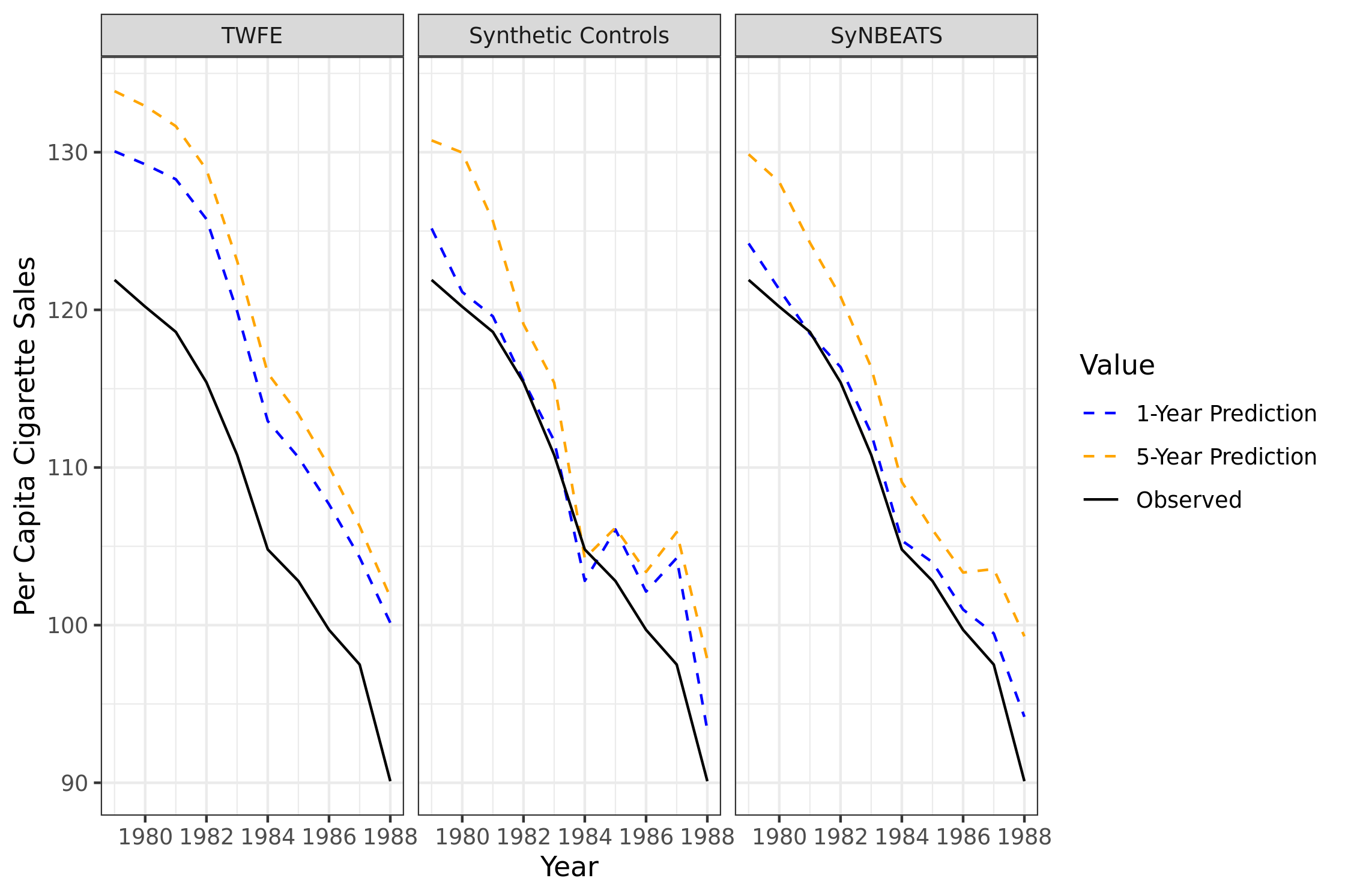}
\caption{{Prop 99 Prediction Errors from Placebo Treated Years}
\\\\\footnotesize The figure compares short-term (year 1 post-treatment) and long-term (year 5 post-treatment) prediction errors obtained from predicting the per capita cigarette sales in California in pseudo-treated years using various estimators. The blue, dashed line depicts the prediction error in the first pseudo-treated year. The orange, dashed line depicts the predictions in the fifth pseudo-treated year. The predictions are formed using data from California from 1970 through the year prior to the first treated year, and from the control states using data from the treated year.}
\label{f:smoking_years}
\end{figure}

As shown in Table \ref{t:traditional}, SyNBEATS outperforms other traditional estimators in their short-term predictions, improving the RMSE by 31\% compared to the second-best alternative (SC). To further facilitate a direct comparison of short- to long-term predictions for each estimator, in Figure \ref{f:smoking_years}, we contrast predictions in the first year after treatment to those obtained in the fifth year after treatment. Unsurprisingly, all estimators perform worse for longer-term predictions, with the performance difference between SyNBEATS and SC closing as well. Focusing on the full five-year window, SyNBEATS slightly outperformed SC.

\subsubsection*{Application}
In this subsection, we apply SyNBEATS and the other estimators we consider to estimate the effect of Proposition 99 on cigarette sales in California. Specifically, we train the model to predict cigarette sales per capita in California during years prior to 1989, and apply the model out-of-sample to predict California cigarette sales per capita in 1989 onward. Our estimated treatment effects correspond to the difference between these predictions and the actual, observed cigarette sales per capita in California during these years. The results of this exercise are displayed in Figure \ref{f:smoking_real_comparison}. Using SyNBEATS, we estimate that Proposition 99 reduced cumulative cigarette sales per capita by a total of 192 packs in the ten years following its passage. Over the same time horizon, the other estimators imply reductions of 207 packs (SC) and 292 packs (TWFE). In relative terms, the SyNBEATS estimate corresponds to a reduction in cigarette sales of 22\% (p=0.03).\footnote{As suggested above, we use randomization inference and assume California was randomly treated. Specifically, our p-value is derived from a placebo test over pseudo-treated units. We estimate a percentage reduction in smoking that is as large as California's in only one of the 38 placebo states we consider.}

\subsection{German Reunification}
The Berlin Wall fell on November 9, 1989. A year later, on October 3, 1990, West Germany and East Germany officially reunited, effectively marking the end of the Cold War. In a canonical study, \citet{abadie2015} apply the SC method to examine the causal effect of the reunification on the economy of West Germany, using 16 other countries as control units. They find that the reunification reduced West Germany's per capita GDP by about 1,600 USD per year in the period of 1990-2003.

Again, we perform two exercises to assess the accuracy of alternative causal inference methods in this setting. First, we iteratively predict the 1990 GDP of each country in the control group. Second, we iteratively predict West Germany's GDP from 1963 to 1989. 
In this setting as well, SyNBEATS consistently outperforms the other methods we consider, both with respect to short-term and long-term predictions. It also outperforms these alternatives with respect to both predicting pseudo-treatment country outcomes in the true treatment year (Figure \ref{f:gdp_states}) and in predicting true treatment country outcomes in pseudo-treatment years (Figure \ref{f:gdp_years}). 

As above, we apply SyNBEATS and the other estimators we  consider to estimate the effect of the German reunification on West Germany's GDP. Specifically, we train the model to predict the GDP in West Germany during years prior to 1990, and apply the model out-of-sample to predict West Germany GDP in 1990 onward. Our estimated treatment effects correspond to the difference between these predictions and the actual, observed GDP in West Germany during the post-treatment years. The results of this exercise are displayed in Figure \ref{f:gdp_real_comparison}. Using SyNBEATS, we estimate that reunification reduced GDP per capita in West Germany by a total of \$17,766 in the 13 years after it occurred. This corresponds to a decrease of 5\% (p=0.05).\footnote{As above, this p-value is derived from a placebo test over untreated units; only one of the 16 placebo countries we consider has an estimated percentage GDP reduction as large as West Germany's.} In contrast, over the same time horizon, SC implies a GDP reduction of \$15,116. The two-way fixed effects estimator suggests an average \textit{increase} in GDP per capita of \$9,577.

\subsection{Predicting Abnormal Returns for Simulated Stock Market Events}
In this exercise, we rely on data used in \citet{baker2020machine} to predict abnormal stock returns for simulated events. This exercise is well suited for assessing the performance of the various estimators for use in financial event study analyses -- i.e., estimates of the causal effect of a shock (such as securities litigation or a merger announcement) on stock prices. The data set includes returns for firms with a share price above \$5 between 2009 and 2019. It also contains 10,000 randomly selected, unique firm-level pseudo-events (i.e., the events do not correspond to anything that would be expected to systematically affect the firms' stock price). For each firm-level event, we use returns data for the 250 trading days prior to the event to predict the returns on the event date. As in \citet{baker2020machine}, for each firm, our pool of control units contains all firms with the same four-digit SIC industry code; if there are fewer than eight such firms, we include peers with the same three-digit SIC industry code.

Using 100 randomly selected pseudo-events from this data, we compare the predictions of each estimator for the stock price on the day after the simulated event (Table \ref{t:traditional}). As in the previous settings, SyNBEATS tends to generate the most accurate predictions, although here the performance gains are more modest. One potential explanation for why SyNBEATS does not improve performance as dramatically in this setting is that there is limited information in prior stock performance that can be used to predict future stock performance. 

\section{Comparison to Recent Methods}\label{sec:recent}

So far, we have provided evidence that SyNBEATS outperforms SC and TWFE, two commonly employed methods by social scientists for causal inference in panel data settings. In this section, we compare SyNBEATS to two recently proposed estimators, matrix completion (MC) and synthetic difference-in-differences (SDID), described in Section \ref{sec:SyNBEATS:ex}.

\subsection{Evaluation Using Empirical Applications}

We begin by comparing the performance of SyNBEATS to MC and SDID using the three empirical applications described in Section \ref{sec:exercises}. To implement MC, we implement the version of the estimator introduced in \citet{mcathey2021}, which has been modified to allow for the estimation of causal effects in panel data settings. 


To implement SDID, we follow \citet{sdid} and their implementation in the associated code release. In particular, we first construct the level-shifted synthetic control with the unit weights described in Section \ref{sec:SyNBEATS:ex}. Next, a weighted DID analysis is performed, where the pre-treatment average value of the treated unit and its level-shifted synthetic control are both given as weighted averages determined by the SDID time weights. The post-treatment average values are just simple averages, and a standard DID analysis is then applied. We note that SDID yields an average treatment effect on the treated unit over the entire post-treatment period of interest. To obtain predictions over varying treatment horizons, we recover the per-period treatment effects by conducting the last DID analysis step on \textit{each} value in the post-treatment period rather than the simple average over all periods. 

To compare SyNBEATS with MC and SDID, we replicate the analyses in the previous section. The results are presented in Table \ref{t:modern}. SyNBEATS dramatically outperforms MC in each analysis we consider with the data sets corresponding to Proposition 99 and the German Reunification. With respect to SDID, the results are more nuanced: SyNBEATS performs better in 6 out of the 8 analyses we consider using these two datasets, but SDID out-performs SyNBEATS in the remaining two. Notably, the performance gains of SyNBEATS compared to SDID are smaller than with the other estimators we consider. Finally, in predicting stock returns, SyNBEATS yields the lowest RMSE, but all three of the estimators yield comparable performance -- again, consistent with the hypothesis that time series forecasts are unlikely to greatly improve the predictive power in this context.

\newpage
\begin{landscape}
\begin{table}[!ht]
\caption{Comparison of Panel Data Methods}
\begin{tabularx}{\hsize}{l|XXXX|XXXX|X}
\toprule
\multicolumn{1}{c|}{Method} & \multicolumn{4}{c|}{Prop 99} & \multicolumn{4}{c|}{German Reunification} & \multicolumn{1}{c}{Stocks} \\ 

& \multicolumn{2}{c}{Units} & \multicolumn{2}{c|}{Years} & \multicolumn{2}{c}{Units} & \multicolumn{2}{c|}{Years} & \\
& Short & Long & Short & Long & Short & Long & Short & Long & \\
\cmidrule(r){1-10}
TWFE & $14.420$ & $15.386$ & $8.369$ & $10.193$ & $2,254.893$ & $2,103.349$ & $797.574$ & $824.017$ & $0.0284$ \\
Synthetic Controls & $7.705$ & $11.152$ & $2.716$ & $4.264$ & $878.390$ & $1,122.949$ & $115.919$ & $305.509$ & $0.0281$ \\ 
SyNBEATS & \boldsymbol{$3.591$} & \boldsymbol{$8.176$} & \boldsymbol{$1.882$} & \boldsymbol{$4.088$} & \boldsymbol{$325.792$} & \boldsymbol{$876.349$} & \boldsymbol{$70.747$} & \boldsymbol{$214.964$} & \boldsymbol{$0.0275$} \\
\bottomrule
\end{tabularx}
\\~\\ {\singlespace \footnotesize Notes: This table compares the performance of the different estimators based on the RMSE. Columns labeled ``Prop 99'', ``German Reunification'' and ``Stocks'' describe our different evaluation data sets. Columns labeled ``Units'' refer to analyses that consider pseudo-treated units and columns labeled ``Years'' refer to analyses that consider pseudo-treated years. Columns labeled ``Short'' refer to one-year predictions and columns labeled ``Long'' refer to five-year predictions.
\label{t:traditional}}
\end{table}
\end{landscape}
\clearpage

\subsection{Evaluation Using Simulations}
Our findings thus far suggest that SyNBEATS consistently outperforms TWFE, SC, and MC. In contrast, the results were more mixed with respect to SDID: SyNBEATS achieved a lower RMSE than SDID for 7 out of the 9 evaluations we considered, whereas SDID yielded lower prediction error for 2 of the 9 evaluations.

To shed additional light on the factors shaping the relative performance of SyNBEATS and SDID, we next conduct a simulation exercise. We consider a data generating process (DGP) via a linear factor model of the form studied in 
\citet{bai2009, gobillonmagnac2016, xu_2017}, adapted to allow for variation in the relative importance of fixed and interactive effects in generating the outcomes (see notes to Appendix Figure \ref{f:sim_pars} for details). 



In our simulations, we try to recover the null effect and vary three parameters, each time comparing the performance of SyNBEATS to SDID. The parameters we vary were selected to capture a range of panel data settings that empirical researchers may confront in practice.

\begin{itemize}
    \item We vary the number of control units from 10 to 100 in increments of 10.
    \item We vary the number of pre-treatment periods from 10 to 100 in increments of 10.
    \item We vary the relative importance of fixed effects vis-a-vis the other components contributing to the outcome $Y$. 
\end{itemize}
We simulate data for each parameter choice 100 times and estimate outcomes over 4 periods. Model performance is evaluated as the average RMSE over all simulations and post-treatment periods. Results are depicted in Appendix Figure \ref{f:sim_pars}. 

As shown in the figure, SyNBEATS consistently outperforms SDID under a variety of parameter choices. However, the relative advantage of SyNBEATS decreases (and even reverses) for shorter pre-treatment periods and fewer controls. This is consistent with the assumption that the greater flexibility in the functional form of SyNBEATS requires more training data to generate reliable estimates. Further, the relative advantage of SyNBEATS decreases as the relative importance of fixed effects increases. As we confirm with additional simulations (Appendix Figure \ref{f:sim_pars_c95}), SDID more consistently outperforms SyNBEATS if the fixed effects strongly dominate the data-generating process. This is consistent with the functional form assumptions of SDID being well-suited for that scenario.

Overall, our results suggest that SyNBEATS appears more accurate if researchers have sufficient training data. However, if training data is sparse, or if the data generating process is known to be dominated by unit fixed effects, SDID may yield lower prediction errors.

\section{Understanding SyNBEATS' Performance}\label{sec:gains}

In this section, we investigate the source of SyNBEATS' strong observed performance relative to alternative estimators. SyNBEATS differs from other estimators in two important ways: (1) its use of both horizontal and vertical information to inform its imputation, and (2) its use of the N-BEATS residual block architecture. But, unlike many common estimators, SyNBEATS is not guaranteed to be consistent; it could be that its performance gains in terms of RMSE stem from providing concentrated, but potentially biased estimates. We explore each of these possibilities in turn.

\subsection{Use of Horizontal and Vertical Information}

To assess the importance of combining both horizontal and vertical information, we compare SyNBEATS to alternative estimators that, although comparable in their architecture, have access to limited information. The first estimator uses a pure forecasting algorithm that does not take into account the information from control states. Instead, it uses data only on pre-treatment outcomes of the treated unit. This approach is exactly the implementation of N-BEATS as originally proposed by \citet{oreshkin2019}. 
The second estimator removes lagged outcomes of the treated unit and instead uses only information from the control states, in a similar way to synthetic controls. To implement it, we adapt the N-BEATS architecture by stripping out the treated unit's lags, translating the problem into one with no time-series dependence. We maintain the same residual block structure, using control unit outcomes to predict the treated unit's outcome in each period, with the assumption that the relationship between units is stable across time.

Appendix Table \ref{t:nocovariates} compares the performance of our baseline SyNBEATS estimator to these variants. Unsurprisingly, SyNBEATS performs the best, but the performance degradation from excluding horizontal information is dramatically larger than from excluding vertical information. This suggests that much of the performance gains in SyNBEATS might be traced back to its ability to efficiently learn the time series structure of the treated unit's outcomes. Without vertical information, SyNBEATS performs similar to or slightly better than TWFE or SC, whereas it performs substantially worse than competing estimators for the longer-term predictions. These results suggest that incorporating the post-treatment information from the control units becomes increasingly important as the prediction horizon grows and the accuracy of the forecast degrades. 

\subsection{Model Architecture}
To investigate the importance of the model architecture upon which SyNBEATS relies, Appendix Table \ref{t:ols} compares SyNBEATS to alternative methods for imputing counterfactual treated unit outcomes based on vertical and horizontal information. SyNBEATS substantially outperforms ``off-the-shelf'' neural network and random forest models. A possible explanation is that SyNBEATS' residual structure allows for efficient mining of the signal without much data, in a similar manner to boosted trees. In a highly overparametrized regime where the number of model parameters far exceeds training data available, SyNBEATS' residual-based architecture allows the model to make full use of the limited data available and converge quickly in a setting where other flexible estimation methods typically suffer. In contrast, the performance gains are present, but less dramatic, when SyNBEATS is compared to a simple linear model. We thus conclude that SyNBEATS' strong observed performance is due to both its model architecture and its ability to use this architecture to exploit information from horizontal and vertical sources efficiently.

\subsection{Consistent Estimation via Bias-Correction}\label{sec:correction}

Finally, the performance gains from SyNBEATS could stem from a sacrifice of consistency in favor of providing concentrated, but biased estimates. Due to potential misspecification arising from flexible estimation, a researcher interested in a consistent estimator cannot simply optimize for mean squared error alone. To explore this idea, we implement a correction procedure in which we estimate and then adjust for the incurred bias.

Recall that SyNBEATS estimates the counterfactual control outcomes $\hat{Y}_{Nt}(0)$ and consequently the treatment effect as: 
\begin{equation*}
    \hat{\tau}^{\text{SyNBEATS}} = \frac{1}{T-T_0} \sum_{t>T_0} \left(Y_{Nt} - \hat{Y}_{Nt}(0)\right)
\end{equation*}

The bias-corrected SyNBEATS estimator employs SyNBEATS twice:first to obtain a treatment effect estimate, and second to obtain a bias estimate for adjustment.  We begin by splitting the pre-treatment period equally. Using the first half of the pre-treatment data, $t=1,\ldots, \lfloor T_0 /2 \rfloor$, as the training set, we train SyNBEATS to construct counterfactuals $\hat{Y}_{Nt,\text{pre}}(0)$ for the second half of the pre-treatment data, $t=\lfloor T_0/2 \rfloor +1, \ldots, T_0$. This yields an estimate of the estimator's bias. We then subtract this estimated bias from the original SyNBEATS (trained on the full pre-treatment period) estimate: 
\begin{equation*}
    \hat{\tau}^{\text{BC}} = \underbrace{\frac{1}{T-T_0} \sum_{t>T_0} \left(Y_{Nt} - \hat{Y}_{Nt}(0)\right)}_{\text{SyNBEATS Estimator}} - \underbrace{\frac{1}{T_0/2} \sum_{t = \lfloor T_0/2 \rfloor +1}^{T_0} \left( Y_{Nt} - \hat{Y}_{Nt,\text{pre}}(0) \right)}_{\text{Bias Correction}}
\end{equation*}
Both the bias incurred by the SyNBEATS estimator and the estimated bias in the second half of the pre-treatment period converge in probability to the same limit. Thus, the difference between the two vanishes asymptotically. 

\citet{viviano} prove that under the following set of canonical event study assumptions, such a procedure results in consistent estimates. Formally, the distribution of outcomes $(Y_{Nt}(0), Y_{1t}, \ldots, Y_{N-1,t})$ is stationary, and the treatment timing is exogenous. Together these ensure that the empirical distributions of the pre- and post-treatment outcomes are invariant across time periods. Similar conditions are implied in the Synthetic Controls literature \cite{abadie2010, sdid, ttest} which treat $T_0$ as deterministic (and hence, exogenous). Next, one imposes standard $\beta$-mixing conditions \cite{viviano}, so that the observations are asymptotically independent. Lastly, the treated unit's potential outcome under treatment can be decomposed as  $Y_{Nt}(1) = \mu_{Nt} + \epsilon_{Nt}$, where $\mu_{Nt}<\infty$ is a possibly non-stationary expectation, and $\epsilon_{Nt}$ is  an additive stationary shock, with $\E[\epsilon_{Nt}] = 0$.

We now assess the importance of consistency guarantees by comparing the performance of this bias-corrected estimator to our baseline (non-bias-corrected) SyNBEATS estimator. We illustrate the empirical effects of bias correction in the California Proposition 99 and West Germany Reunification datasets. We revisit the setting of our exercise with pseudo-treated, where we restrict the data to time periods before the actual treatment occurs. In the California Proposition 99 dataset, we iteratively predict the per-capita cigarette sales from 1975 to 1988; in the West Germany Reunification dataset, we iteratively predict West Germany's GDP from 1965 to 1989.\footnote{This time window differs slightly from our pseudo-treated years exercise above (1963-1989), as we split the pre-treatment period in half in the bias-adjustment method but still require a non-zero number of observations for training in the bias estimation.} The error distribution of the SyNBEATS estimator and the bias-adjusted SyNBEATS estimator is shown in Figure \ref{f:bias-corrected}.



For this bias-correction procedure to be meaningful, the gains in bias must be large relative to the increase in variance. Across both datasets, we see that applying the the bias-correction shifts the mean error closer to zero, with a larger (percentage) change in the Prop 99 setting. In the Prop 99 setting, this reduction in bias comes at the cost of a substantial increase in variance, but in the West Germany setting, applying the bias correction leads to both slightly lower bias and slightly lower variance (Table \ref{t:bc}).\footnote{One explanation for this is finite sample noise, coupled with the already strong performance of the original SyNBEATS estimator. As our estimator minimizes mean squared error, the very modest improvement in bias and marginal reduction in variance may be a result of overall lower mean squared error due to randomness in the training process. }

Researchers interested in implementing SyNBEATS will have to decide the extent to which asymptotic unbiasedness is important for their application. For example, if a researcher further decides to tune the SyNBEATS model via time-series cross-validation techniques, overfitting and regularization biases may be exacerbated, in which case a bias-correction may be valuable. Overall, our results in this subsection suggest that combining SyNBEATS with bias-correction techniques can reduce the bias of estimated treatment effects in empirical applications, but in some cases this can come at a substantial reduction in precision.

\section{Conclusion}\label{sec:conclusion}

This paper introduced the SyNBEATS algorithm as a tool for causal inference in panel data settings. In the applications we considered, SyNBEATS consistently yielded lower prediction errors compared to commonly employed estimators (SC and TWFE), as well as comparable or stronger performance than more recently developed approaches (MC and SDID). Supplemental analyses suggest that model architecture and the use of both vertical and horizontal information contribute to SyNBEATS' strong performance. 
Overall, compared to alternative estimators, SyNBEATS may yield more accurate estimates of the causal effect of policies of interest in a range of realistic panel-data settings.

Our focus has been on developing SyNBEATS and assessing the accuracy of its performance on real data; we have not focused on quantifying the uncertainty of the resulting estimates. In settings where researchers require more conventional standard errors, the approaches that have been proposed for SC-type estimators can be readily adapted to SyNBEATS. For example, the researcher can apply SyNBEATS to placebo treatments obtained by permuting the treatment year or treatment unit \cite{abadie2010,shaikh2021randomization}. Alternatively, instead of modeling the assignment mechanism directly, one can quantify uncertainty in model prediction using conformal inference methods to obtain finite sample coverage guarantees \cite{conformalinf}. Lastly,  \citet{viviano} show how one can exploit stationarity and use the block-bootstrap procedure \cite{politis1994stationary} to construct valid $t$-statistics under a sharp null.

\newpage 

\bibliographystyle{aer} 
\bibliography{library} 

\clearpage
\setcounter{page}{1}
\renewcommand{\thepage}{Online Appendix-\arabic{page}}
\appendix
\section*{Online Appendix (Not For Publication)}
\setcounter{figure}{0} \renewcommand{\thefigure}{A.\arabic{figure}}
\setcounter{table}{0} \renewcommand{\thetable}{A.\arabic{table}}

\begin{figure}[!ht]
    \centering
    \includegraphics[scale=0.175]{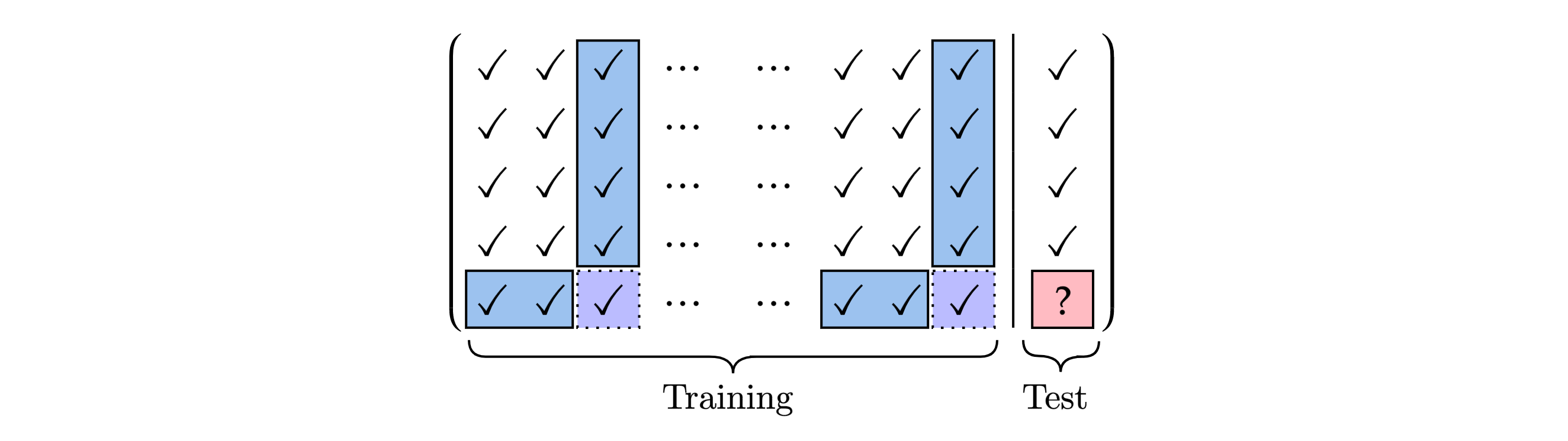}
    \caption{{SyNBEATS on $\boldsymbol{Y(0)}$}
    \\\\\footnotesize Our model casts the dataset imputation as a supervised learning problem. In particular we mask historical outcomes (purple) and learn to predict them with lagged outcomes and contemporaneous controls (blue). We then apply this model out of sample to the true missing data (red). 
    } 
    \label{fig:syn}
\end{figure}

Traditional estimators exploit the structure of the data either via a ``vertical'' \cite{abadie2003, abadie2010} or ``horizontal'' \cite{imbens_wooldridge} regression, as shown in panels B and C of Figure \ref{fig:intro_problem}. The blue regions in each image indicate which subsets of the data are being used to find the respective weights for each method. Our method, SyNBEATS, aims to exploit both patterns simultaneously by casting this imputation problem as a supervised learning task, shown in Figure \ref{fig:syn}. Note that the number of lags is a hyperparameter.

\begin{figure}[!ht]
\centering
\includegraphics[width=.8\textwidth] {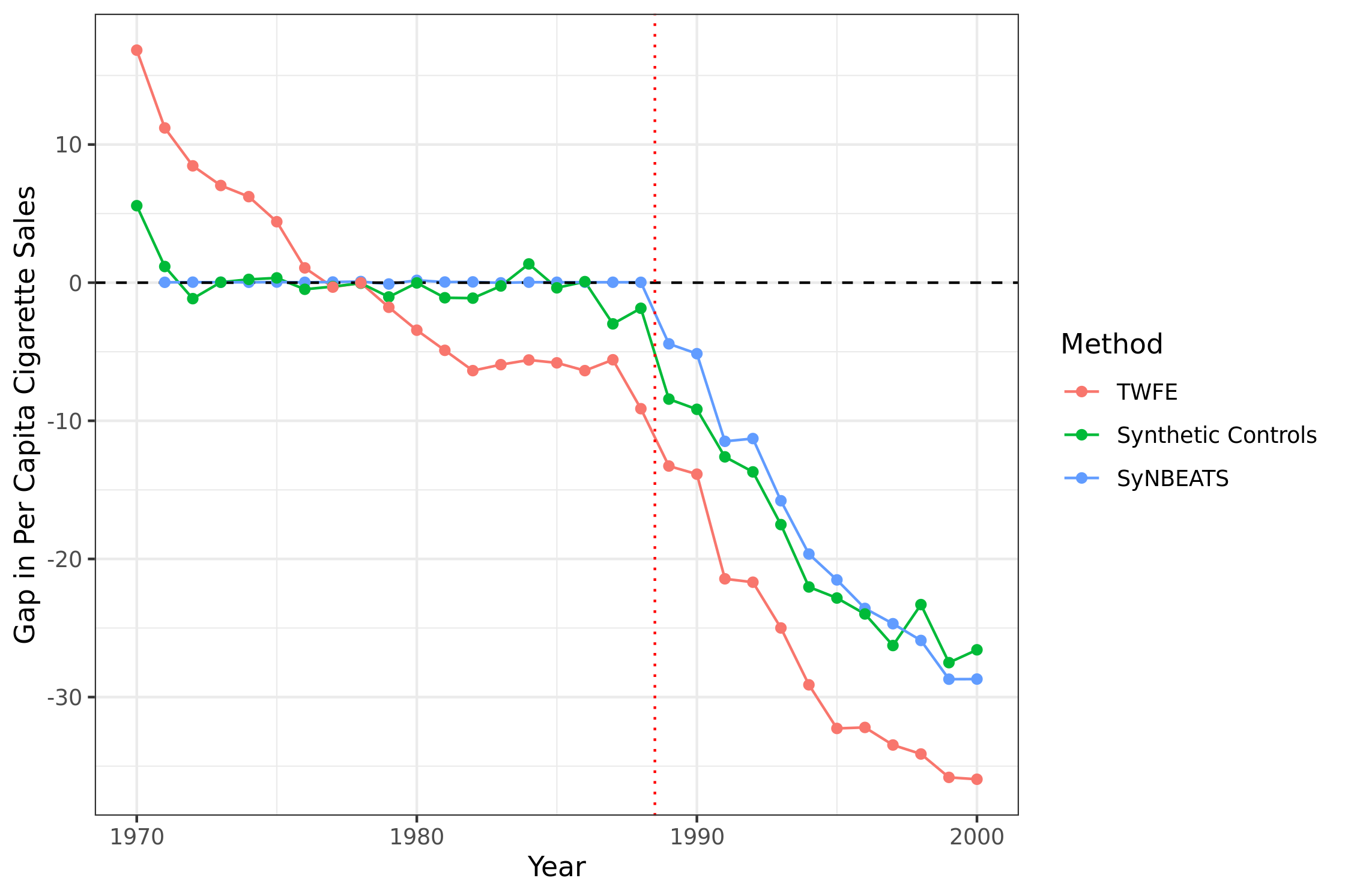}
\caption{{Effect of Prop 99: Predicted Versus Observed Cigarette Sales}
\\\\\footnotesize This figure compares the estimated effect of Proposition 99 on cigarette sales from 1970 to 2000 across the different estimators. The predictions are formed using data from California in 1970-1988 and for the control states from 1970-2000. The red dashed line represent the treatment year.}
\label{f:smoking_real_comparison}
\end{figure}

\begin{figure} 
\centering
\subfloat[1-Year Predictions]{\includegraphics[width=.8\textwidth] {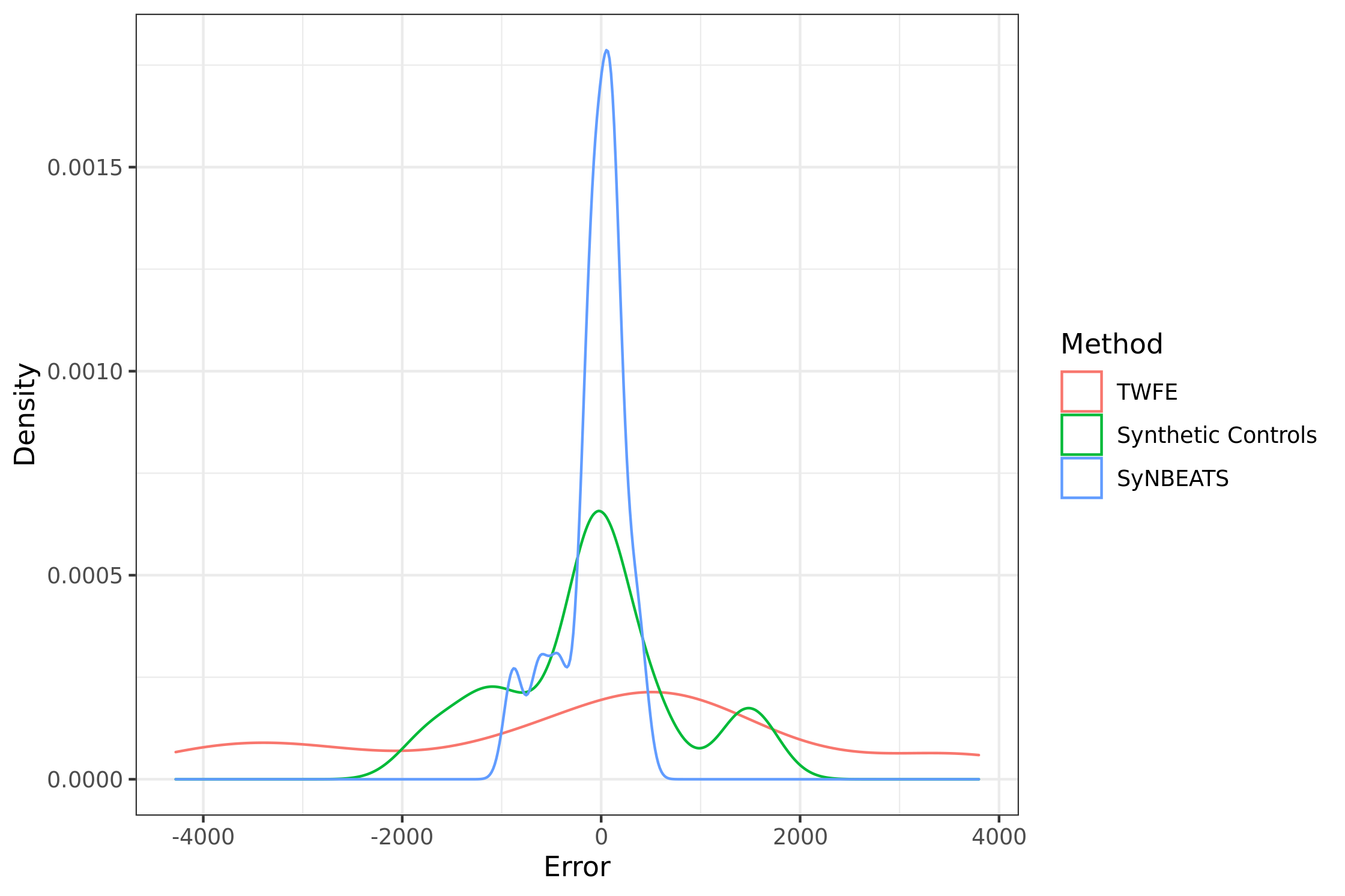}} \\
  \subfloat[5-Year Predictions]{\includegraphics[width=.8\textwidth] {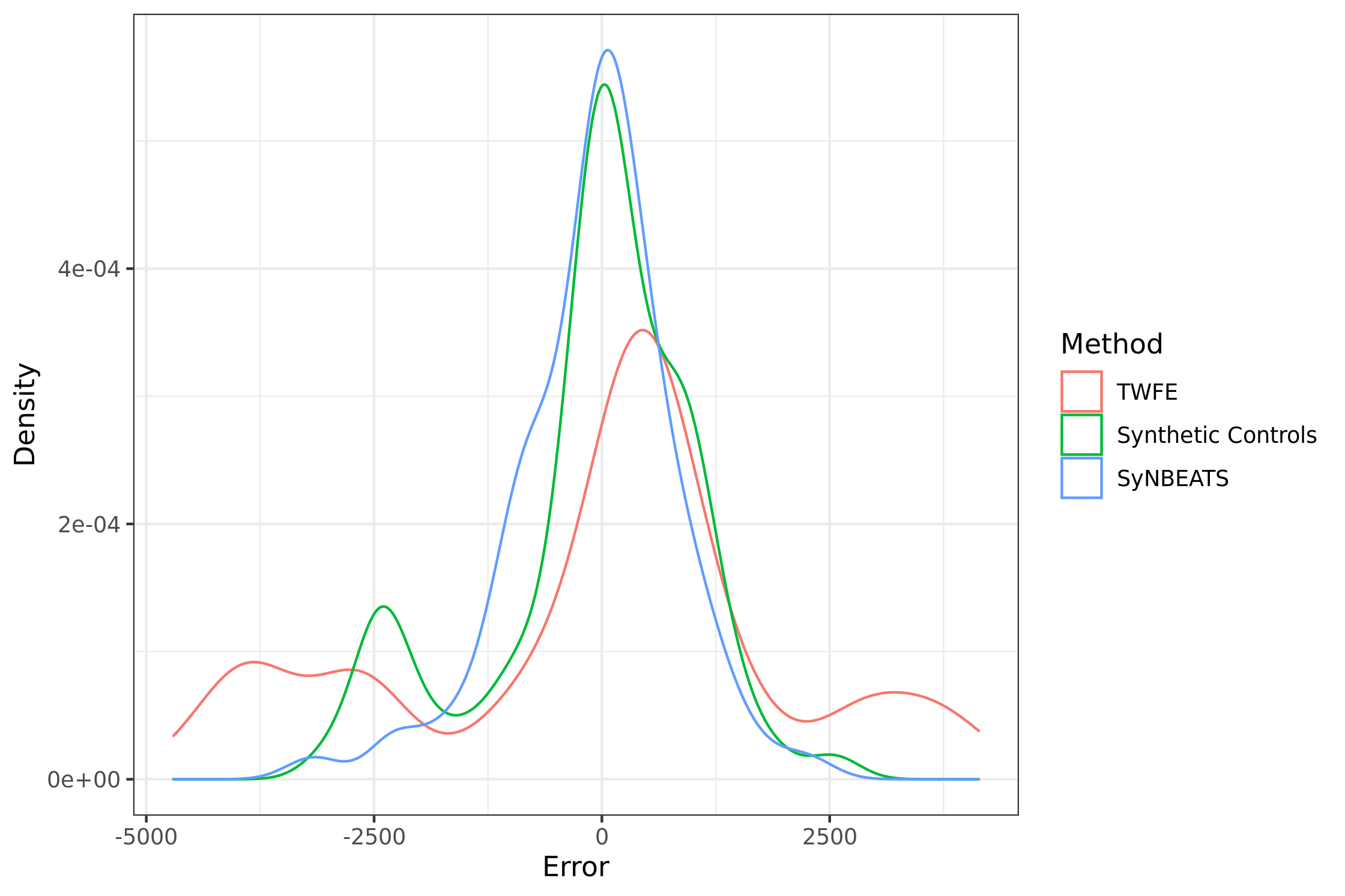}}
\caption{{German Reunification Prediction Errors from Placebo Treated Countries}
\\\\\footnotesize The figure shows the distribution of prediction errors for pseudo-treated states, by prediction method, for the German reunification data set. In Panel A, the outcome being predicted is the pseudo-treated country's GDP in 1990 ($N=16$); the model is trained on data from the pseudo-treated state from 1960-1989 and from the other control states for 1990. In Panel B, the outcome being predicted is the pseudo-treated country's GDP in years 1986-1990; the model is trained on data from the pseudo-treated country from 1960-1985 and from the other control countries for the year being predicted.}
\label{f:gdp_states}
\end{figure}

\begin{figure}[!ht]
\centering
\includegraphics[width=.8\textwidth] {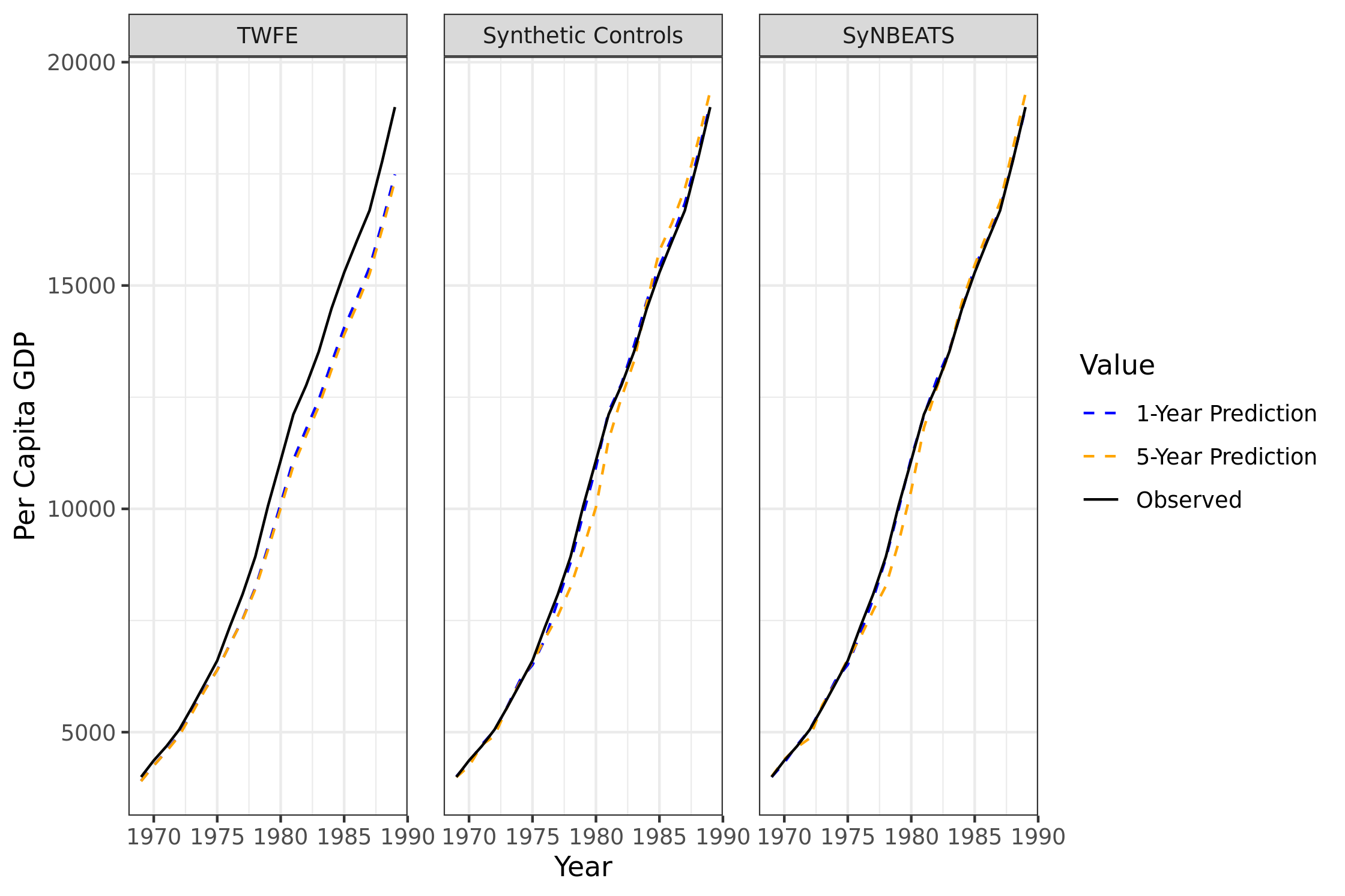}
\caption{{German Reunification Prediction Errors from Placebo Treated Years}
\\\\\footnotesize The figure compares short-term (year 1 post-treatment) and long-term (year 5 post-treatment) prediction errors obtained from predicting the GDP in West Germany in pseudo-treated years using various estimators. The blue, dashed line depicts the prediction error in the first pseudo-treated year. The orange, dashed line depicts the predictions in the fifth pseudo-treated year. The predictions are formed using data from West Germany from 1960 through the year prior to the first treated year, and from the control countries using data from the treated year.}
\label{f:gdp_years}
\end{figure}

\begin{figure} 
\centering
\includegraphics[width=.8\textwidth] {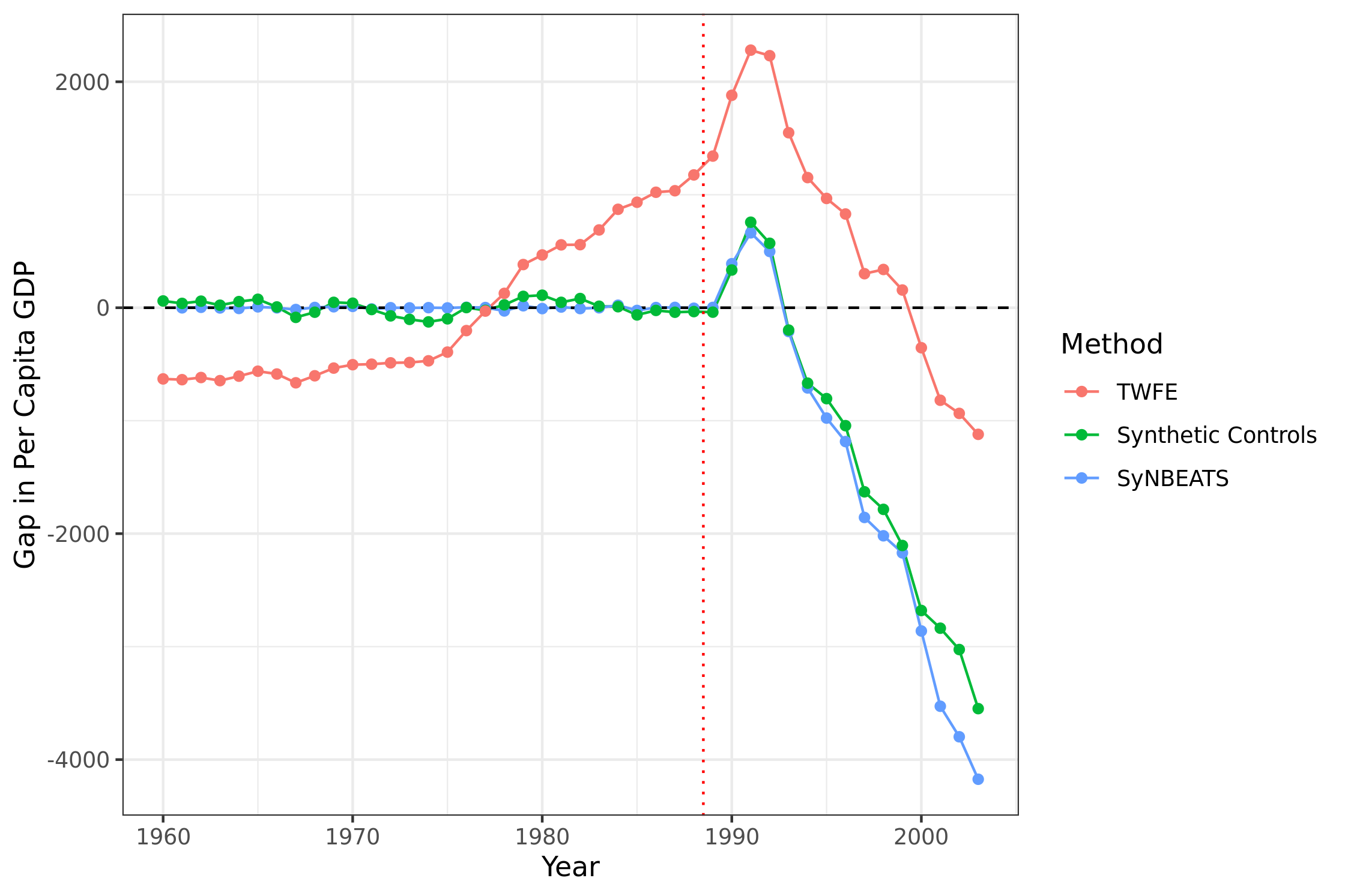}
\caption{{Effect of German Reunification: Predicted Versus Observed GDP}
\\\\\footnotesize This figure compares the estimated effect of the German reunification on the GDP of West Germany from 1990 to 2003 across the different estimators. The predictions are formed using data from West Germany in 1962--1989 and for the control states from 1962--2002. The red dashed line represent the treatment year.}
\label{f:gdp_real_comparison}
\end{figure}

\begin{figure} 
\centering
\includegraphics[width=.8\textwidth] {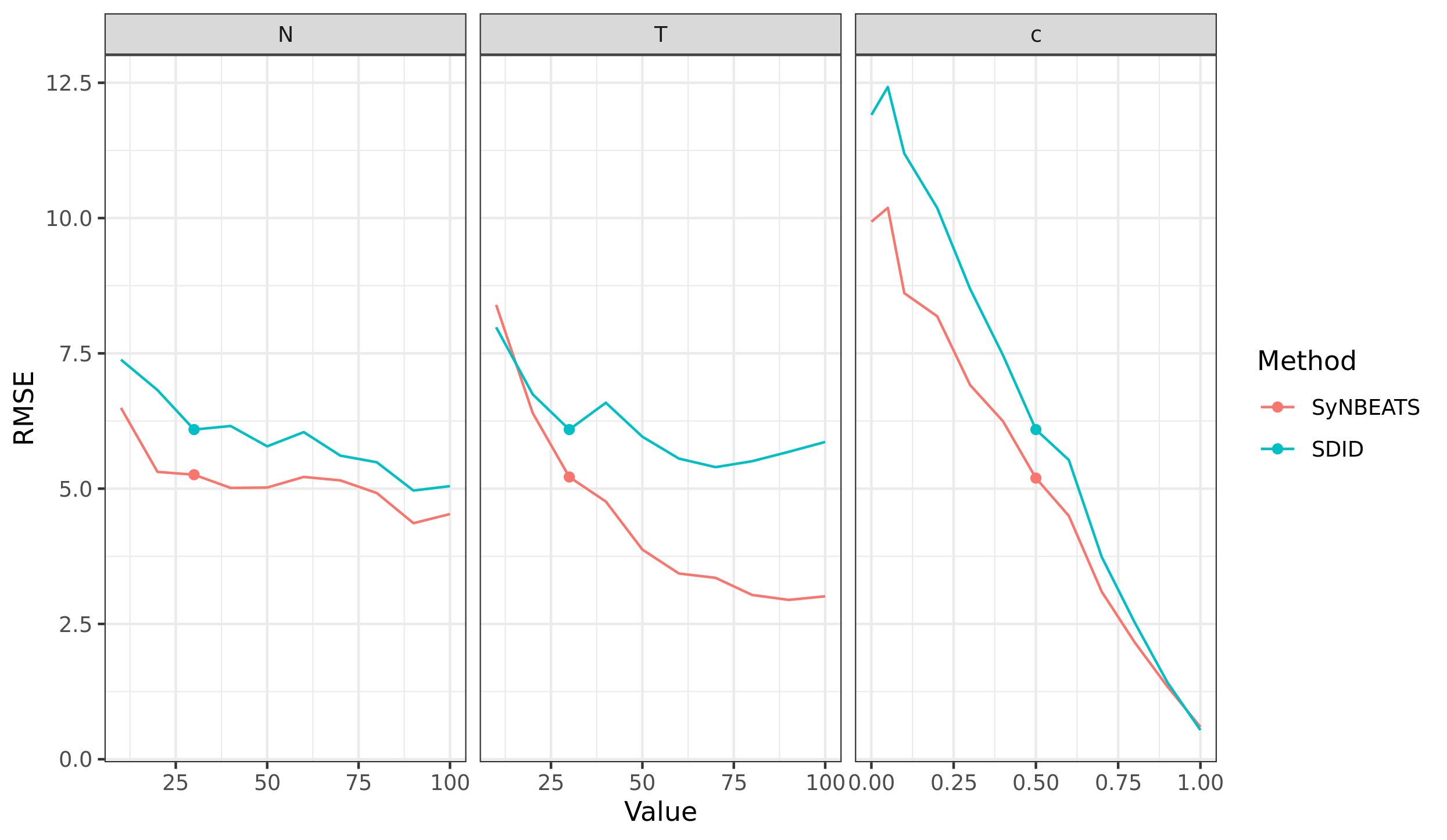}
\caption{{Simulation Comparison of SDID and SyNBEATS}
\\\\\footnotesize The figure compares root mean squared error (RMSE) of predictions generated by SyNBEATS and Synthetic Difference-in-Differences (SDID) under a range of simulated panel data settings. The DGP is given by: $Y_{it} = c*\left(\alpha_i + \xi_t\right) + (1-c)*\left(x_{it,1} \cdot 1 + x_{it,2} \cdot 3 + \lambda_i'f_t + 5 \right) + \epsilon_{it}$, where $f_t = (f_{1t},\ldots,f_{pt})'$ and $\lambda_i = (\lambda_{i1}, \ldots \lambda_{ip})'$ are $p$-dimensional time-varying factors and unit-specific factor loadings, respectively, and $\epsilon_{it} \sim N(0,1)$. Factors and time fixed effects are drawn similarly: $f_{1t},\ldots, f_{pt}, \xi_t \sim N(0,1)$. Factor loadings and unit fixed effects are drawn uniformly with zero mean and unit variance: $\lambda_{i1}, \ldots \lambda_{ip}, \alpha_i \sim U[-\sqrt{3},\sqrt{3}]$. In this model specification, the regressors are positively correlated with factors, loadings, and their product. For $k=1,2$, regressors are given by $x_{it,k} = 1 + \lambda_i' f_t + \lambda_{i1} + \lambda_{i2} + f_{1t} + f_{2t} + \eta_{it,k}$, where $\eta_{it,1},\eta_{it,2} \sim N(0,1)$. The baseline simulation assumes $30$ control units, $30$ pre-treatment time periods, and equal importance of fixed effects and interactive components (i.e., $c = 0.5$). For ease of reference, the red and blue dots depict the performance under the baseline specification. The panel labeled `N' varies the number of control units (from 0 to 100); the panel labeled `T' varies the number of pre-treatment time periods (from 0 to 100); and the panel labeled `$c$' varies the relative importance of fixed effects to interactive effects, with higher values indicating greater importance of the fixed effects component of the data generating process. Loosely speaking, a higher $c$ corresponds to a lower complexity DGP. The plotted RMSE corresponds to the average per-period RMSE over 4 post-treatment time periods from 100 draws of a data generating process with the specified parameters.
}
\label{f:sim_pars}
\end{figure}

\begin{figure} 
\centering
\includegraphics[width=.8\textwidth] {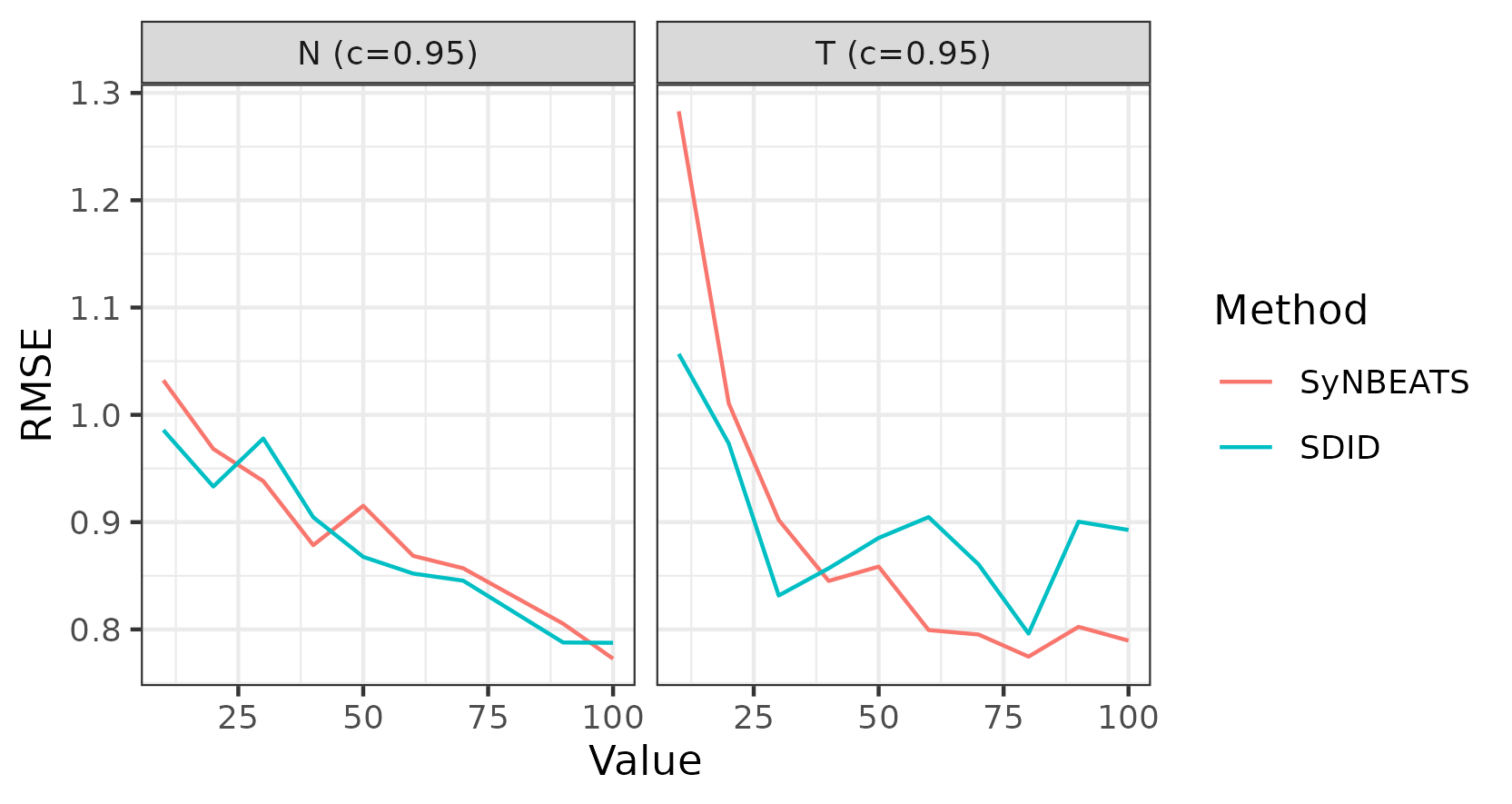}
\caption{{Simulation Comparison of SDID and SyNBEATS for simpler DGPs}
\\\\\footnotesize The figure compares root mean squared error (RMSE) of predictions generated by SyNBEATS and Synthetic Difference-in-Differences (SDID) under a range of simulated panel data settings. The DGP is given by: $Y_{it} = c*\left(\alpha_i + \xi_t\right) + (1-c)*\left(x_{it,1} \cdot 1 + x_{it,2} \cdot 3 + \lambda_i'f_t + 5 \right) + \epsilon_{it}$, where $f_t = (f_{1t},\ldots,f_{pt})'$ and $\lambda_i = (\lambda_{i1}, \ldots \lambda_{ip})'$ are $p$-dimensional time-varying factors and unit-specific factor loadings, respectively, and $\epsilon_{it} \sim N(0,1)$. Factors and time fixed effects are drawn similarly: $f_{1t},\ldots, f_{pt}, \xi_t \sim N(0,1)$. Factor loadings and unit fixed effects are drawn uniformly with zero mean and unit variance: $\lambda_{i1}, \ldots \lambda_{ip}, \alpha_i \sim U[-\sqrt{3},\sqrt{3}]$. In this model specification, the regressors are positively correlated with factors, loadings, and their product. For $k=1,2$, regressors are given by $x_{it,k} = 1 + \lambda_i' f_t + \lambda_{i1} + \lambda_{i2} + f_{1t} + f_{2t} + \eta_{it,k}$, where $\eta_{it,1},\eta_{it,2} \sim N(0,1)$. In all simulations, $c=0.95$, indicating a high level of relative importance of fixed effects to interactive effects. The panel labeled `N' varies the number of control units (from 0 to 100); the panel labeled `T' varies the number of pre-treatment time periods (from 0 to 100). The plotted RMSE corresponds to the average per-period RMSE over 4 post-treatment time periods from 100 draws of a data generating process with the specified parameters.
}
\label{f:sim_pars_c95}
\end{figure}

\begin{table}[!ht]
\caption{German Reunification Analysis with Placebo Treated Countries}
\begin{tabularx}{\textwidth}{XXXX}
\toprule
Method & RMSE & MAPE & Best  \\
\cmidrule(r){1-4}
\multicolumn{4}{c}{\em 1 Year Predictions}\\
\cmidrule(r){1-4}
TWFE & $2,254.893$ & $11.498$ & $0$ \\ 
Synthetic Controls & $878.390$ & $3.884$ & $0.312$ \\ 
SyNBEATS & \boldsymbol{$325.792$} & \boldsymbol{$1.423$} & \boldsymbol{$0.688$} \\ 

\cmidrule(r){1-4}
\multicolumn{4}{c}{\em 5 Year Predictions}\\
\cmidrule(r){1-4}
TWFE & $2,103.349$ & $11.713$ & $0.188$ \\ 
Synthetic Controls & $1,122.949$ & $6.049$ & \boldsymbol{$0.438$} \\
SyNBEATS & \boldsymbol{$876.349$} & \boldsymbol{$4.330$} & $0.375$ \\
\bottomrule
\end{tabularx}
\\~\\ {\singlespace \footnotesize Notes: This table summarizes performance of the estimators we consider at predicting the GDP for each pseudo-treated country, using the German reunification data set. Each model is trained on data from the pseudo-treated country and from the other control states for the year(s) being predicted. In Panel A, root mean-squared error (RMSE) and mean absolute percentage error (MAPE) are calculated from the distribution of prediction errors across pseudo-treated countries for the year following treatment (1990). In Panel B, RMSE and MAPE are calculated from the distribution of average annual prediction errors across pseudo-treated countries for the five-year period following treatment (1986-1990). The column ``Best'' reports the share of pseudo-treated countries for which the specified estimation method yields the lowest prediction error over the specified time horizon (i.e., either one year or average over the five-year period).

\label{t:gdp_states}}
\end{table}

\begin{table}
\caption{German Reunification Analysis with Placebo Treated Years}
\begin{tabularx}{\textwidth}{XXXX}
\toprule
Method & RMSE & MAPE & Best  \\
\cmidrule(r){1-4}
\multicolumn{4}{c}{\em 1 Year Predictions}\\
\cmidrule(r){1-4}
TWFE & $797.574$ & $5.159$ & $0.120$ \\
Synthetic Controls & $115.919$ & $1.180$ & $0.160$ \\ 
SyNBEATS & \boldsymbol{$70.747$} & \boldsymbol{$0.840$} & \boldsymbol{$0.720$} \\
\cmidrule(r){1-4}
\multicolumn{4}{c}{\em 5 Year Predictions}\\
\cmidrule(r){1-4}
TWFE & $824.017$ & $5.679$ & $0.095$ \\ 
Synthetic Controls & $305.509$ & $2.258$ & $0.190$ \\ 
SyNBEATS & \boldsymbol{$214.964$} & \boldsymbol{$1.686$} & \boldsymbol{$0.714$} \\ 
\bottomrule
\end{tabularx}
\\~\\ {\singlespace \footnotesize Notes: This table summarizes performance of the estimators we consider at predicting the GDP in West Germany for each pseudo-treated year, using the German reunification data set. Each model is trained on data from West Germany from 1960 through the year prior to the first pseudo-treated year, and from the control countries using data from the pseudo-treated year. In Panel A, root mean-squared error (RMSE) and mean absolute percentage error (MAPE) are calculated from the distribution of prediction errors across pseudo-treated years (1963--1989). In Panel B, RMSE and MAPE are calculated from the distribution of average annual prediction errors across for each five-year pseudo-treatment period (i.e., 1963--1967 through 1985--1989). The column ``Best'' reports the share of pseudo-treated years for which the specified estimation method yields the lowest prediction error over the specified time horizon (i.e., either one year or average over the five-year period).
\label{t:gdp_years}}
\end{table}

\newpage
\begin{landscape}
\begin{table}[!ht]
\caption{Comparison of Modern Panel Data Estimators}
\begin{tabularx}{\hsize}{l|XXXX|XXXX|X}
\toprule
\multicolumn{1}{c|}{Method} & \multicolumn{4}{c|}{Prop 99} & \multicolumn{4}{c|}{German Reunification} & \multicolumn{1}{c}{Stocks} \\ 

& \multicolumn{2}{c}{Units} & \multicolumn{2}{c|}{Years} & \multicolumn{2}{c}{Units} & \multicolumn{2}{c|}{Years} & \\
& Short & Long & Short & Long & Short & Long & Short & Long & \\
\cmidrule(r){1-10}
Matrix Completion & $7.173$ & $13.446$ & $4.422$ & $9.083$ & $905.532$ & $1,305.247$ & $207.155$ & $670.672$ & 0.0279 \\ 
SDID & $3.742$ & $8.745$ & \boldsymbol{$1.769$}  & $4.747$ & $382.673$ & \boldsymbol{$656.548$} & $71.623$ & $219.838$ & $0.0283$ \\ 
SyNBEATS & \boldsymbol{$3.591$} & \boldsymbol{$8.176$} & $1.882$ & \boldsymbol{$4.088$} & \boldsymbol{$325.792$} & $876.349$ & \boldsymbol{$70.747$} & \boldsymbol{$214.964$} & \boldsymbol{$0.0275$} \\
\bottomrule
\end{tabularx}
\\~\\ {\singlespace \footnotesize Notes: This table compares the performance of modern estimators (Matrix Completion and Synthetic Difference-in-differences) based on the RMSE. Columns labeled ``Prop 99'', ``German Reunification'' and ``Stocks'' describe our different evaluation data sets. Columns labeled ``Units'' refer to analyses that consider pseudo-treated units and columns labeled ``Years'' refer to analyses that consider pseudo-trated years. Columns labeled ``Short'' refer to one-year predictions and columns labeled ``Long'' refer to five-year predictions. 
\label{t:modern}}
\end{table}
\end{landscape}
\clearpage

\newpage


\begin{landscape}
\begin{table}[!ht]
\caption{Comparison of Baseline SyNBEATS to Restricted Models}
\begin{tabularx}{\hsize}{l|XXXX|XXXX|X}
\toprule
\multicolumn{1}{c|}{Method} & \multicolumn{4}{c|}{Prop 99} & \multicolumn{4}{c|}{German Reunification} & \multicolumn{1}{c}{Stocks} \\ 

& \multicolumn{2}{c}{Units} & \multicolumn{2}{c|}{Years} & \multicolumn{2}{c}{Units} & \multicolumn{2}{c|}{Years} & \\
& Short & Long & Short & Long & Short & Long & Short & Long & \\
\cmidrule(r){1-10}
No Covariates & $5.27$ & $12.79$ & $3.10$ & $9.59$ & $367.33$ & $1467.53$ & $237.52$ & $944.74$ & $0.031$ \\ 
No Lags & 44.22 & 61.06 & 23.64 & 31.58 & 2766.34
& 3780.69 & 13954.64 &15452.41 & 0.041 \\
Baseline & \boldsymbol{$3.59$} & \boldsymbol{$8.18$} & \boldsymbol{$1.88$} & \boldsymbol{$4.09$} & \boldsymbol{$325.79$} & \boldsymbol{$876.35$} & \boldsymbol{$70.78$} & \boldsymbol{$214.96$} & \boldsymbol{$0.028$} \\ 
\bottomrule
\end{tabularx}
\\~\\ {\singlespace \footnotesize Notes: This table replicates the analyses reported in Table \ref{t:traditional} and Appendix Table \ref{t:modern} for the comparison of SyNBEATS with restricted versions of itself, in order to better understand where its gains arise. The ``No Covariates'' estimator differs from the baseline SyNBEATS estimator in that its predictions for the treated unit are based entirely on prior observations from the treated unit, and do not include values of the control units. The ``No Lags'' estimator differs from the baseline SyNBEATS estimator in that its predictions for the treated unit are based entirely on contemporaneous values of the control units.
\label{t:nocovariates}}
\end{table}
\end{landscape}
\clearpage

 \begin{table}
 \caption{Comparison of SyNBEATS with Alternative Model Architectures} 
\begin{tabularx}{\textwidth}{X XXXXX} 
\toprule
{} &  OLS &  NN  &  RF & {SyNBEATS} \\
\midrule
1 Lags & 	5.51 &	30.97 & 8.28  & {3.59}\\ 
2 Lags &  6.00	& 25.06	& 8.68  & {3.41}\\ 
3 Lags &  5.23 & 	27.95 & 	8.33  & {3.29}\\ 
4 Lags & 	5.28 & 	30.06	& 8.43  & {3.67}\\ 
5 Lags & 	4.25 & 	27.16 & 	8.59  & {3.36}\\ 
\bottomrule
\label{t:simplemodel}
\end{tabularx}
 \\~\\{\footnotesize\singlespace Notes: This table summarizes pseudo-treated states RMSE on the Proposition 99 dataset, comparing SyNBEATS to alternative estimators that make use of both vertical and horizontal information. Each estimator includes as a feature the contemporaneous outcomes for control units, as well as the specified number of lags of the units being predicted. In the ``OLS'' column  we use an unrestricted linear regression to predict the treated outcome based on both the contemporaneous control and lagged treated outcomes. NN denotes a fully-connected neural network using a three-layer perceptron with a hidden layer of 100 neurons. RF denotes a random forest with 100 trees.
 }
\label{t:ols}
\end{table}

\begin{figure}
    \centering
    \includegraphics[width=0.8\textwidth]{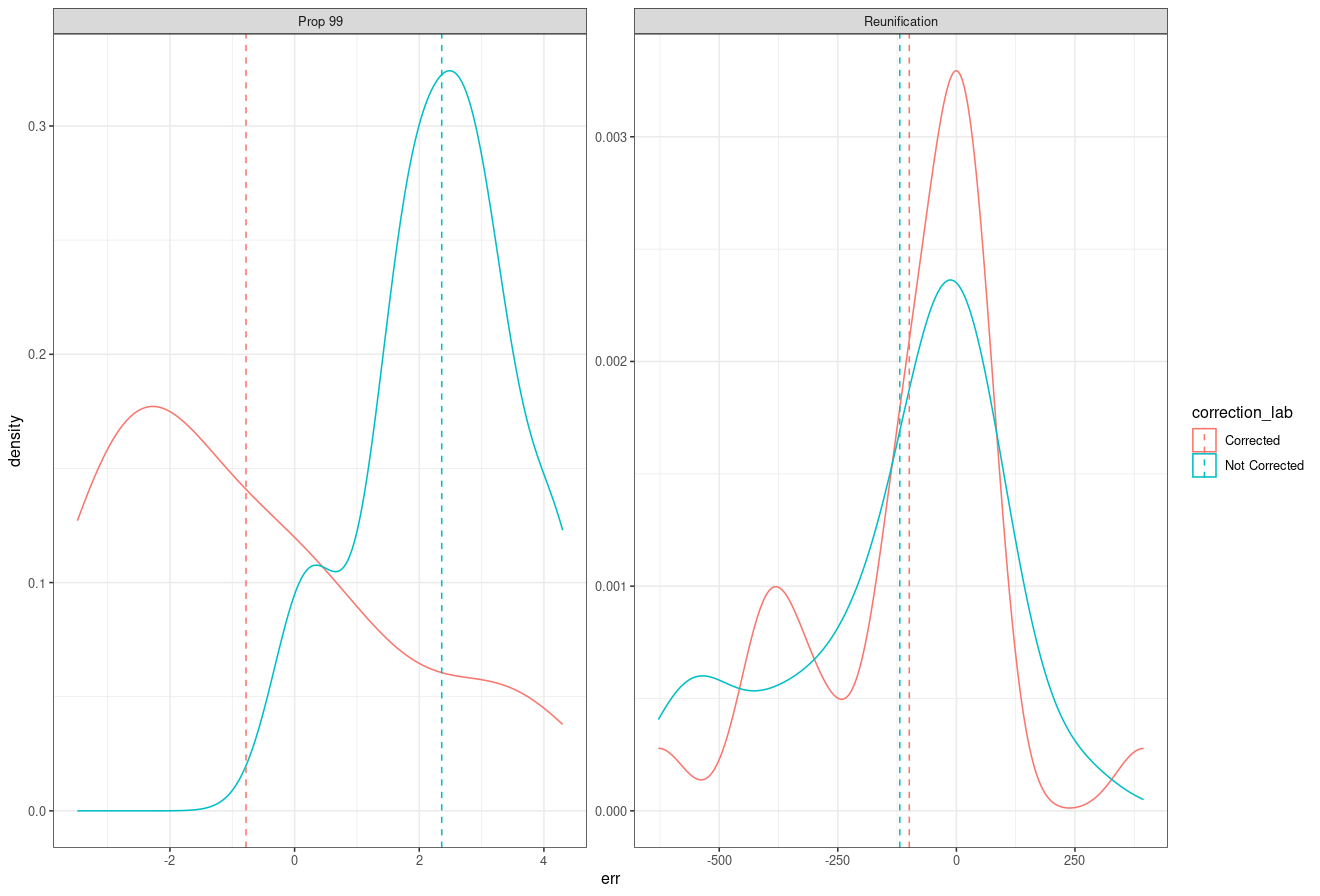}
    \caption{{Prop 99 and Reunification Prediction Errors, SyNBEATS vs. Bias-corrected SyNBEATS.}
    \\\\\footnotesize The figure shows the distribution of prediction errors for our pseudo-treated years exercise, bias-corrected and bias-uncorrected, for both the Proposition 99 and Reunification datasets. In the left panel, the outcome  predicted is California's per-capita cigarette sales tax; in the right panel, the outcome predicted is West Germany's GDP. The dashed line represents the mean of the prediction errors.}\label{f:bias-corrected}
\end{figure}

\begin{table}
\caption{Comparison of SyNBEATS with and without bias-correction} 
\begin{tabularx}{\textwidth}{ccccc} 
\toprule
\multicolumn{1}{c}{} & \multicolumn{2}{c|}{Mean Error} & \multicolumn{2}{c}{Variance} \\ 
\multicolumn{1}{c}{} &  \multicolumn{1}{c}{Baseline} & \multicolumn{1}{c|}{Bias-corrected}  &  \multicolumn{1}{c}{Baseline} & \multicolumn{1}{c}{Bias-corrected} \\
\midrule
California Proposition 99 & 2.36 & -0.78 & 1.51 & 5.18 \\  
    West Germany Reunification & -119 & -99.3 & 47815 & 43783 \\
\bottomrule
\end{tabularx}
 \\~\\{\footnotesize\singlespace Notes: This table summarizes the mean and variance of the error distributions for the pseudo-treated years exercise using SyNBEATS (baseline) and its bias-corrected version, as described in the text.
 }
\label{t:bc}
\end{table}

\end{document}